\documentclass[lettersize,journal]{IEEEtran}
\usepackage{authblk}
\usepackage{amsmath,amsfonts}
\usepackage{amssymb}
\usepackage{array}
\usepackage[caption=false,font=normalsize,labelfont=sf,textfont=sf]{subfig}
\usepackage{textcomp}
\usepackage{stfloats}
\usepackage{url}
\usepackage{verbatim}
\usepackage{graphicx}
\usepackage{cite}
\usepackage[normalem]{ulem}
\def\BibTeX{{\rm B\kern-.05em{\sc i\kern-.025em b}\kern-.08em
    T\kern-.1667em\lower.7ex\hbox{E}\kern-.125emX}}
\usepackage{balance}
\usepackage{algorithmic}
\usepackage[utf8]{inputenc}
\usepackage{float}
\usepackage[T1]{fontenc}
\usepackage{braket}
\usepackage{graphicx}
\usepackage{amsthm}
\usepackage{placeins}
\usepackage{amsfonts}
\usepackage{tabularx}
\usepackage{makecell}
\usepackage{physics}
\usepackage{enumitem}
\usepackage{hyperref}
\usepackage{placeins}
\usepackage{xcolor}
\usepackage{booktabs}
\usepackage{amsmath}
\usepackage{quantikz}
\usepackage[english]{babel}
\usepackage{underscore}

\usepackage{algorithm}

\usepackage[symbol]{footmisc}

\newcommand{\CZ}{\mathrm{CZ}}
\newcommand{\CX}{\mathrm{CX}}

\newtheorem{definition}{Definition}

\newcounter{protocol}
\newlist{protsteps}{enumerate}{1}                
\setlist[protsteps]{label=\textbf{Step~\arabic*:}, 
                    leftmargin=*,                
                    topsep=0.01\baselineskip,     
                    itemsep=0.01\baselineskip}
\newenvironment{protocol}[1][]{%
  \refstepcounter{protocol}
  \par\addvspace{0.3\baselineskip}
  \noindent\textbf{Protocol~\theprotocol}\ifx\relax#1\relax\else: #1\fi%
  \par\nobreak\smallskip
  \begin{protsteps}}
 {\end{protsteps}\par\addvspace{\baselineskip}}

\usepackage{titlesec}
\titlespacing*{\section} {0pt}{1ex}{1ex}
\titlespacing*{\subsection} {0pt}{1ex}{1ex}
\titlespacing*{\paragraph} {0pt}{1ex}{1ex}

\usepackage[singlelinecheck=false,font=small,skip=5pt]{caption}
\DeclareMathAlphabet{\mathcal}{OMS}{cmsy}{m}{n}
\SetMathAlphabet{\mathcal}{bold}{OMS}{cmsy}{b}{n}

\usepackage{xpatch} 
\makeatletter
\xpatchcmd\@collaboration@present{(}{\medskip}{}{}
\xpatchcmd\@collaboration@present{)}{}{}{}
\makeatother



\begin{document}
\title{Piecemaker: a resource-efficient entanglement distribution protocol}
    \author[1]{Luise Prielinger\textsuperscript{*}\thanks{\textsuperscript{*}l.p.prielinger@tudelft.nl}}
    \author[2]{Kenneth Goodenough}
    \author[2]{Guus Avis}
    \author[2]{Stefan Krastanov}
    \author[2]{Don Towsley}
    \author[2]{Gayane Vardoyan}

    \affil[1]{QuTech, Delft University of Technology, Delft, The Netherlands.}
    \affil[1]{EEMCS, Delft University of Technology, Delft, The Netherlands.}
    \affil[2]{MCICS, University of Massachusetts, Amherst, USA.}
    
\maketitle
\begin{abstract} 
We introduce multipartite entanglement distribution protocols that use a quantum switch to deliver stabilizer states to a number of remote end users.
As in existing schemes, the first step in our protocols involves Bell pair generation between the switch and each end user. However,
unlike existing schemes that wait for all Bell pairs to be established before distributing the desired state---for example, via a projective measurement---our approach stores only a minimal subset of Bell pairs while processing every subsequent Bell pair immediately. In doing so, our protocols reduce the average Bell pair storage time compared to existing schemes,
resulting in less cumulative noise as a direct consequence.
On the theoretical side, our protocol design is grounded in the structure of vertex covers in graph states up to local complementation.
Through a comprehensive numerical evaluation, we compare the fidelities of delivered states with those of a baseline scheme, for state sizes up to $n=50$ qubits. Simulations also show that our protocols can achieve the critical fidelity threshold of $\frac{1}{2}$ for multipartite entanglement in a wider range of depolarization rates and success probabilities of Bell-pair generation. Overall, our protocols always achieve an equal or higher fidelity of the distributed state, and can reduce infidelity by up to 45\%.
\end{abstract}


\section{Introduction}
In many distributed quantum applications, entanglement forms a core resource. In particular, the entanglement of stabilizer states has been found to be a resource vital for sensing, secret-sharing and computation~\cite{PhysRevA.52.R2493,bartolucci2021fusionbased,Kitaev2003,Briegel,RevModPhys.83.33,Azuma2015, Damian2020}. Distributing such states is a non-trivial task, and inevitably leads to noise being imparted onto the state. 

\begin{figure}
    \centering
    \includegraphics[width=0.8\linewidth]{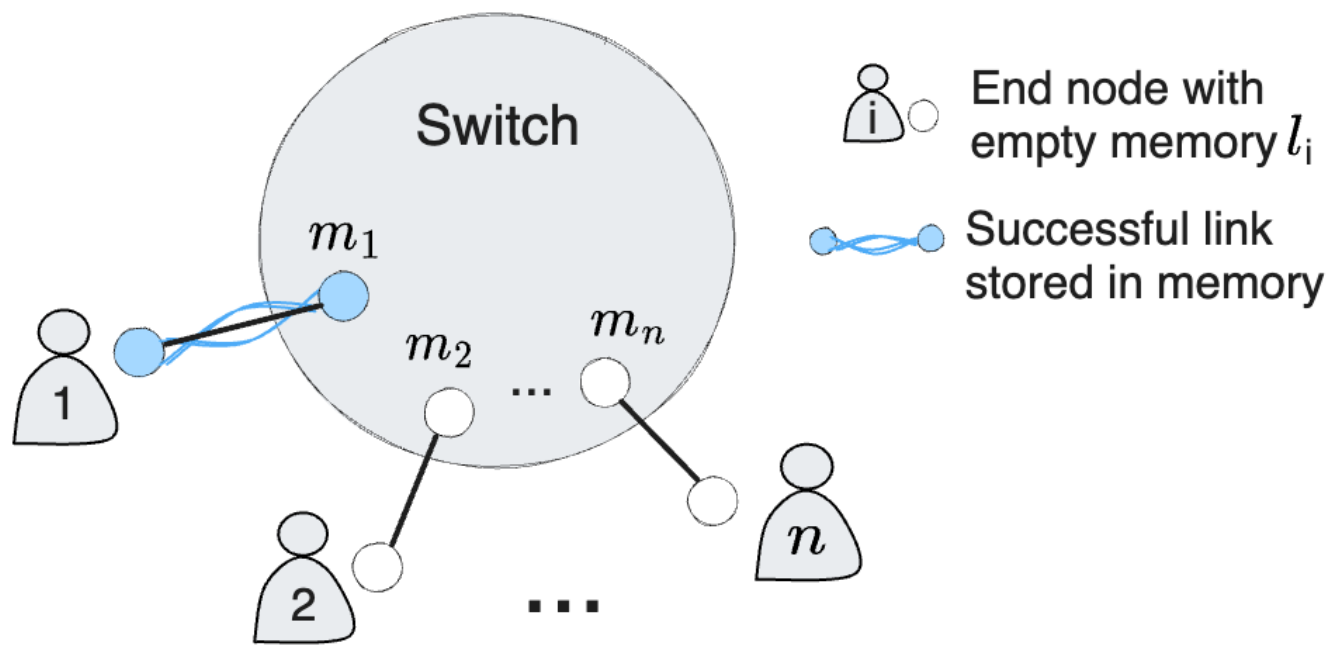}
    \caption{Sketch of a central switch node connected to $n$ end nodes.
    The central node has quantum memories $\set{m_1, \dots, m_n}$,
    while each end node $i$ can hold one qubit $l_i$.}
    \label{fig:model}
\end{figure}

Quantum entanglement switches have recently been intensely studied for their multipartite entanglement distribution capabilities~\cite{Fischer2021,avis2023analysis,vardoyan2019stochastic,bugalho2023distributing,nain2022analysis,vardoyan2020exact,promponas2024full,bhatti2025distributing, sen2023multipartite,gauthier2023control,shimizu2025simple}. Figure \ref{fig:model} shows a simplified model of such a quantum entanglement switch. We build on the case where a multipartite entangled state is distributed by first creating the target state locally at a central node and then teleporting it to the end nodes~\cite{avis2023analysis}. Here, the central node, termed \emph{Factory}, firstly attempts to create a shared Bell pair---which we refer to as an \emph{entangled link}, or simply \emph{link} for brevity---with each of the end nodes. After all the links have been established, any desired target state can be created at the Factory node, which is then transmitted qubit-by-qubit to the end nodes using quantum teleportation. In general, these entangled links are created at different times, leading to two problems: first, they need to be stored until all of them have been established, resulting in the degradation of these bipartite entangled states and thus also of the final delivered state, as qubits stored in noisy memory accumulate errors over time. Second, the fact that the memories are occupied until the full state has been distributed prevents the memories from being used for other applications. In this study we introduce protocols for distributing multipartite states that mitigate the above problems. 
An important example is the Greenberger-Horne-Zeilinger (GHZ) state~\cite{greenberger1990bell}, which plays a crucial role in many quantum network applications~\cite{murta2020quantum, hillery1999quantum}. For GHZ states, 
our protocol only stores the first link until the protocol completes, while every subsequent link is processed immediately, occupying memory only for the time required to establish it and reducing the probability that storage induces an error.
Our protocol is therefore more resilient to memory decoherence, and we provide empirical evidence for this in Section~\ref{sec:numerical}.

Furthermore, while many of the aforementioned works focus on the generation and distribution of GHZ states only~\cite{ shimizu2025simple, vardoyan2019stochastic}, we extend our protocols to general stabilizer states. In essence, while one can create a large GHZ state by fusing smaller GHZ states or Bell pairs through simple two-qubit operations and Pauli measurements~\cite{Ghaderibaneh2023Generation}, the same holds true for generating and distributing stabilizer states~\cite{sen2023multipartite}. That is, we assemble Bell pairs  in a resource efficient manner using simple two-qubit operations to create a large stabilizer state. We achieve the latter using a strategy similar to the one we outlined for GHZ states. First, the switch waits until a sufficient set of links is present---i.e., a set that meets the minimal requirements for the protocol to proceed, see the next paragraph. Afterwards, every subsequent link is processed immediately, releasing its associated memory.

We determine the above-mentioned minimal sets of links that need to be present at the switch, by utilizing an important notion in stabilizer states, namely \emph{local Clifford (LC) equivalence}~\cite{van2004efficient, van2004graphical}. In essence, any stabilizer state can be transformed into any given LC-equivalent stabilizer state with a series of local Clifford operations. We can first create an intermediate LC-equivalent state for which only a known subset of links needs to be stored in memory, and then restore the desired final state using local Clifford operations before its delivery to the end nodes. In particular, these minimal sets of links correspond to vertex covers up to local complementation in graph states, see Section~\ref{sec:Graphprot} for more details.

In current quantum-networking setups, memory qubits are a scarce resource that remains difficult to control~\cite{ruf2021quantum}. Our network end nodes are therefore assumed to have only simple capabilities with storage capacity for one qubit, the ability to perform heralded entanglement generation with the switch node, and the ability to apply single-qubit Pauli gates and measurements. The switch has storage capacity for up to $n$ qubits, as well as the ability to execute two-qubit gates. Note that the Factory as described in \cite{avis2023analysis} utilizes $2n$ qubits at the central node. However, the Factory protocol can be adapted to require only $n$ qubits by replacing the teleportation step by a projective measurement in a basis defined by the target state. In contrast, our protocol explicitly describes how any stabilizer state can be distributed using only Clifford gates and measurements in the standard basis.

Our contributions can be summarized as follows:
\begin{itemize}[noitemsep,topsep=0pt]
    \item We introduce protocols that  are more noise resilient in distributing stabilizer states, compared to the ones found in existing literature for similar quantum network setups;
    \item We conduct a comprehensive numerical evaluation for system sizes up to $n=50$ end nodes, where Bell pairs are created with equal probability, revealing that 
    performance improvements can be significant. For example, when distributing GHZ states, the infidelity can be lowered by up to $45\%$ on average when using our protocol over the Factory protocol~\cite{avis2023analysis}. Further, we characterize the behavior of our protocol when it comes to a given threshold fidelity. For example, we find that our protocol always reduces the minimal required hardware parameters to achieve the critical fidelity threshold of $\frac{1}{2}$ in all system sizes;
    \item In addition, we analyze our protocol when the probability to create a Bell pair between the switch and an end node varies across end nodes. Here, we find that in many parameter regimes, our protocol is less affected by this inhomogeneity than the baseline protocol~\cite{avis2023analysis}.
\end{itemize}

The rest of the paper is organized as follows.
First, we introduce the switch and other relevant details of the quantum networking model in Section \ref{sec:qnmodel}. Next, we explain in detail how our protocols operate. In Section \ref{sec:GHZprot} we introduce a protocol that generates an $n$-GHZ state from Bell pairs piece by piece. We term this protocol the \emph{GHZ Piecemaker} protocol. Subsequently, in Section \ref{sec:Graphprot}, we generalize the piece-wise distribution approach to arbitrary stabilizer states; which is why we term this protocol the \emph{(general) Piecemaker} protocol. In the last section, Section \ref{sec:numerical}, we numerically compare the performance of our protocols with the teleportation protocol introduced in~\cite{avis2023analysis}. 

\section{Quantum Network Model} \label{sec:qnmodel}
We consider a star-topology quantum network as drawn in Figure \ref{fig:model}, where a central switch connects to each of $n\geq 2$ end nodes through classical and quantum channels. Each end node $i\in \set{1,2,\dots,n}$ possesses one memory qubit, which we refer to as $l_i$. The switch has a storage capacity of $n$ memory qubits $\{m_1, \dots, m_n\}$. In order for an end node to share a Bell pair $\ket{\Phi^+_i}= (\ket{00} + \ket{11})/\sqrt{2}$~\cite{einstein1935can} with the switch,  
the node $i\leftrightarrow$ switch quantum channel
continuously attempts to generate entanglement until success.
We assume that entanglement generation is a Bernoulli process that succeeds with probability $p^i_{\mathrm{link}}$ for node $i$.
Each attempt is assumed independent of the other attempts so that the number of attempts until success follows a geometric distribution $\mathrm{Geom}(p^i_{\mathrm{link}})$.

A successfully generated Bell pair between the switch and a node $i$ is stored in quantum memory: one half at the node's memory $l_i$ and the other half within the switch memory $m_i$. 
The outcome of each entanglement generation attempt is heralded to both parties via the classical channel connecting them.
Each attempt takes a constant duration $\Delta t$ which includes heralding time.
Our network thus evolves at discrete time intervals of $\Delta t$, which we call \emph{rounds}.
Our model of heralded entanglement generation is compatible with schemes such as~\cite{bernien2013heralded, hermans2023entangling}, where the nodes use so-called \emph{communication} qubits at each end of the quantum link to create an entangled state. For example, in nitrogen-vacancy centers~\cite{rozpkedek2019near, pompili2021realization}, electron spins serve as communication qubits, which send out photons that travel through the quantum link to an optical midpoint station. When detection at this midpoint station is successful, the outcome is sent back to the nodes. With this outcome, the nodes learn that they have established a link between their electron spins. At last, they move the state from the communication qubits to their memory qubits, completing a successful round of heralded entanglement generation.

The longer a quantum state has to reside in memory, the more susceptible it becomes to environmental disturbances~\cite{nielsen2010quantum}. Since depolarizing noise is considered a worst-case scenario~\cite{king2003capacity}, we treat each memory as being subject to a depolarizing channel
\begin{equation}
  \mathcal{E}_{\mathrm{depol}}(\rho)\ =\ 
      \bigl(1-p_{\mathrm{depol}} \bigr)\ \rho
      +p_{\mathrm{depol}}\frac{I}{2},
  \label{eq:depol-channel}
\end{equation} 
where ${\rho\in\mathbb{C}^{2\times2}}$ is a qubit's density operator and the probability that an error occurs during a fixed time interval $\Delta t$ is
\begin{equation}
  p_{\mathrm{depol}} :=1 - e^{-\Delta t/\tau},
  \label{eq:depol-prob}
\end{equation}
which we use as the single-shot depolarization probability (see more details Methods \ref{methods:simulation}). The coherence time $\tau$ of a quantum memory depends on the specific experimental setup and can vary between milliseconds and seconds~\cite{pompili2021realization, bradley2022robust}. For simplicity, we assume all quantum memories in the quantum network to be identical with the same coherence time.

We assume the switch is equipped with a quantum processor that can execute controlled-$Z$ ($\CZ$) and controlled-$X$ ($\CX$) gates, single-qubit Clifford gates, as well as Pauli measurements. The end nodes can perform only single-qubit Clifford gates and Pauli measurements. 

Importantly, we make the simplifying assumption that all gates and measurements are noiseless and instantaneous. This isolates the effect of memory decoherence, so that we can focus on evaluating only the infidelity caused by imperfect quantum storage. 
For the majority of protocols we consider, the number of gates equals that of the Factory scheme, such that omitting gate noise does not bias the comparison. However, when we study the protocol in which we use single-qubit Clifford operations, additional gates and thus additional noise is introduced. The amount of noise highly depends on the stabilizer state being generated. The evaluation of different stabilizer states---taking into consideration the gates necessary in the corresponding LC-equivalent graphs---deserves its own careful study, which is beyond the scope of this work.

\section{Background and related work}
A variety of methods exist for distributing multipartite entangled states using a quantum switch in a star-shaped network. An alternative setup to the one we study is a purely photonic scheme~\cite{caprara2019high, wang2009schemes, gauthier2023control}. Here, the switch node does not have quantum memories. Instead, incoming photons are optically interfered and measured, resulting in the projection of the target state onto the end nodes. One benefit of such a scheme is that without memory there is also no memory decoherence. However, a successful projection of the state onto the end nodes is highly probabilistic, with maximum success probability of $1/2^n$~\cite{calsamiglia2001maximum}. Further, in order to generate a multipartite state, all photons need to arrive simultaneously, making the memory-less scheme highly sensitive to photon losses. 

Other studies investigate setups where end nodes have more sophisticated capabilities, e.g., the \emph{2-switch} scheme in~\cite{avis2023analysis}, where end nodes hold two qubits and can execute multi-qubit operations, or assume larger memory buffer, see e.g.,~\cite{vardoyan2023capacity}. While these systems have additional  capabilities, they are also more difficult to realize. In our work, the end nodes are assumed to be as simple as possible, holding only one memory qubit that can be manipulated with $X$ or $Z$ Pauli gates. 

To the best of our knowledge, the Factory switch studied in~\cite{avis2023analysis} is the most similar to ours. One difference is that we assume quantum gates and measurements to succeed deterministically, while the prior study~\cite{avis2023analysis} also analyzes the case where Bell state measurements are probabilistic. In this work, we conduct extensive simulations comparing the performance of our protocol to that of the Factory protocol introduced in~\cite{avis2023analysis} using deterministic Bell state measurements.

\section{GHZ Piecemaker Protocol} 
\begin{figure}
    \centering
    \includegraphics[width=0.45\textwidth]{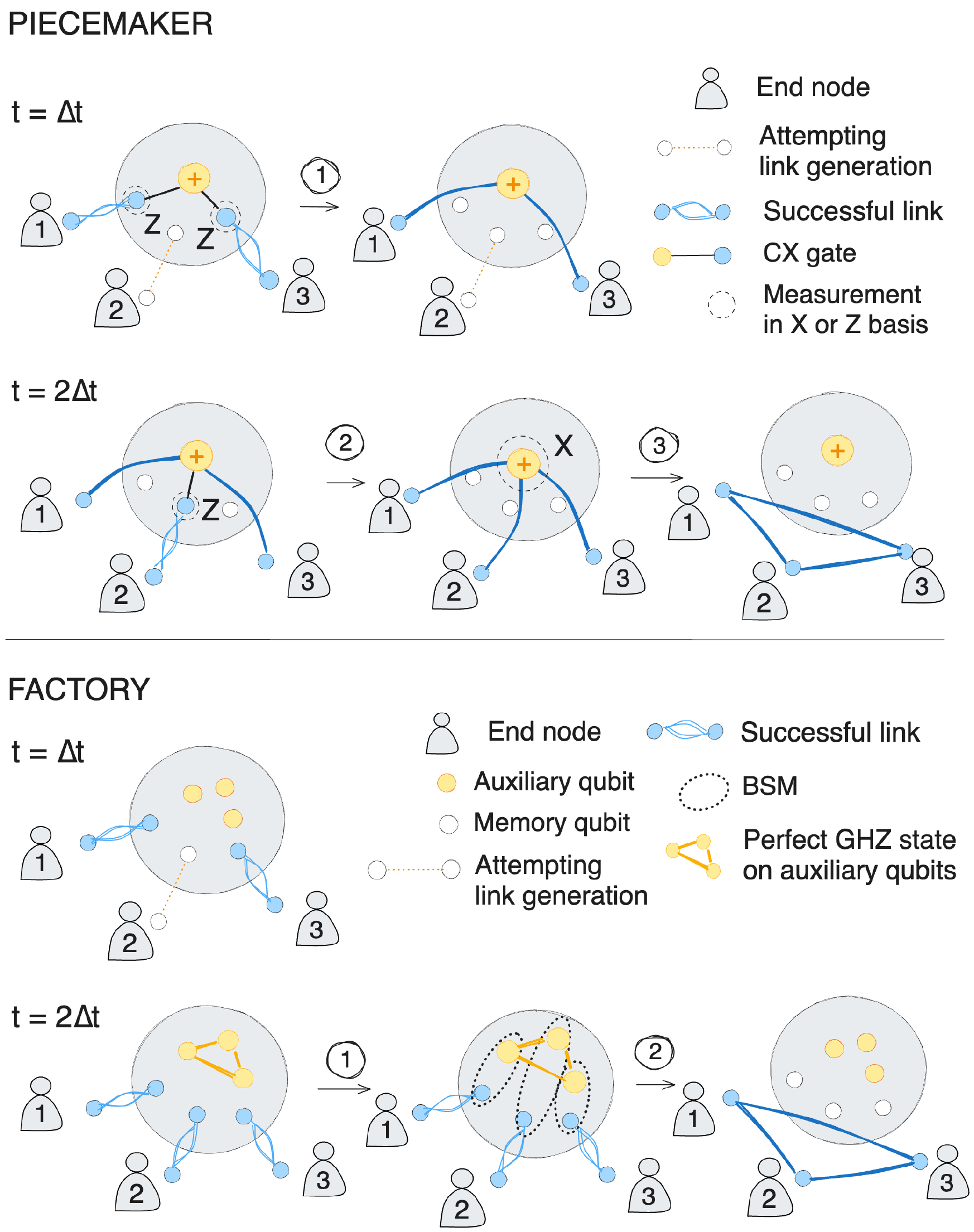}
    \caption{Example instances of the Piecemaker and the Factory schemes. Piecemaker: (1) Once an end node creates a link with the switch, the switch applies a $\CX$ gate between the Piecemaker qubit and that node’s qubit. (2) At $t = 2\Delta t$, all three links have been created and the switch measures the Piecemaker qubit in the $X$ basis. (3) This projects the state onto the end nodes. Factory: (1) Once all links are created, the factory switch prepares a perfect GHZ state on $n$ auxiliary memory qubits and (2) teleports the state to the end nodes using Bell state measurements (BSMs). 
    Note that in both schemes only link generation induces time delay; operations and measurements are assumed to be instantaneous.}
    \label{fig:sketchGHZ}
\end{figure}
\label{sec:GHZprot}
The goal of this protocol is to distribute an $n$-GHZ state piece by piece to $n$ end nodes as efficiently as possible. It achieves this by a simple gate operation that \emph{fuses} entangled links created with the end nodes. Generally, gate-based fusion refers to using unitary gates, such as $\CX$ or $\CZ$, to combine quantum states, often with post-selection or measurement~\cite{nielsen2010quantum, de2020protocols}. We note that the memory-based operations we refer to as fusions are not to be confused with fusing procedures in all-photonic schemes, typically referred to as Type I and Type II fusion~\cite{browne2005resource}.

\begin{definition}[Gate-based fusion] \label{def:fusion}
Let $\set{a_1, b_1, b_2}$ be a set of qubit memories. Perform the following operations:
\begin{enumerate}
    \item Apply a $\CX$ with control $a_1$ and target $b_1$.
    \item Measure $b_1$ in the computational (Z) basis, recording outcome $s\in\{0,1\}$.
    \item If $s=1$, apply a Pauli $X$ gate (correction) to $b_2$, otherwise no correction is needed.
\end{enumerate}
If $a_1$ is part of a GHZ state and $(b_1,b_2)$ hold a Bell pair, this will effectively add $b_2$ to the existing GHZ state.
\end{definition}
Based on this operation, we build our protocol: the switch holds a memory register $M = \set{m_1, m_2, \dots, m_{n+1}}$, where all memories except the last---$m_{n+1}$, which we will refer to as \emph{Piecemaker} qubit---are used by the end nodes to create links. We introduce the Piecemaker qubit solely for explanatory convenience, and explain at the end of this section why the protocol can as well operate without it. The scheme proceeds as described in Protocol~\ref{prot:GHZ}.
\begin{protocol}[GHZ Piecemaker]\label{prot:GHZ}
    \item All end nodes in parallel engage in heralded Bell pair generation.
    \item Whenever a link, say $i$, succeeds, the following actions are performed. If this is the first link to succeed, the Piecemaker qubit $m_{n+1}$ is initialized in the plus state $\ket + = \tfrac 1 {\sqrt 2} (\ket 0 + \ket 1)$. Then the switch fuses the link with the existing state using the set of qubit memories $\set{m_{n+1}, m_i, l_i}$ defined in Def. \ref{def:fusion}, with $l_i$ being the memory of an end node as .
    \item When all links have succeeded (i.e., the previous step has been completed for $i = 1, 2, ..., n$), the switch measures the Piecemaker qubit in the $X$ basis and sends the correction outcome to one of the end nodes for correction.
\end{protocol}

These steps establish an $n$-GHZ state 
\begin{align}
    \ket{\textrm{GHZ}}_n = \frac{1}{\sqrt{2}}(\ket{0_1..0_n} + \ket{1_1..1_n})
\end{align}
shared by the end nodes. We sketch the protocol in Figure~\ref{fig:sketchGHZ}. In contrast to the Factory protocol~\cite{avis2023analysis}, where all links are created before the target state is generated, the Piecemaker protocol allows for immediate piece-wise fusion. This means that at any given time during execution, fewer qubits are involved in the system's multipartite entangled state on average. Take for example the three end-node scenario in Figure~\ref{fig:sketchGHZ}, the Piecemaker occupies one memory qubit at the switch while the Factory occupies two memory qubits. We expect that involving fewer noisy qubits on average will lead to an overall higher quality of the state at the end of the protocol, which we numerically confirm in Section~\ref{sec:numerical}.

We now explain why this protocol can be in principle be carried out without an additional qubit. After the first Bell pair is fused with the Piecemaker qubit, the latter and the end node once again share the same Bell pair. Because this fusion incurs no cost, we could just as well fuse the qubit that carries the first link with every subsequent link.

\section{General Piecemaker Protocol} \label{sec:Graphprot}
\begin{figure*}[ht]
    \centering
    \includegraphics[width=0.9\textwidth]{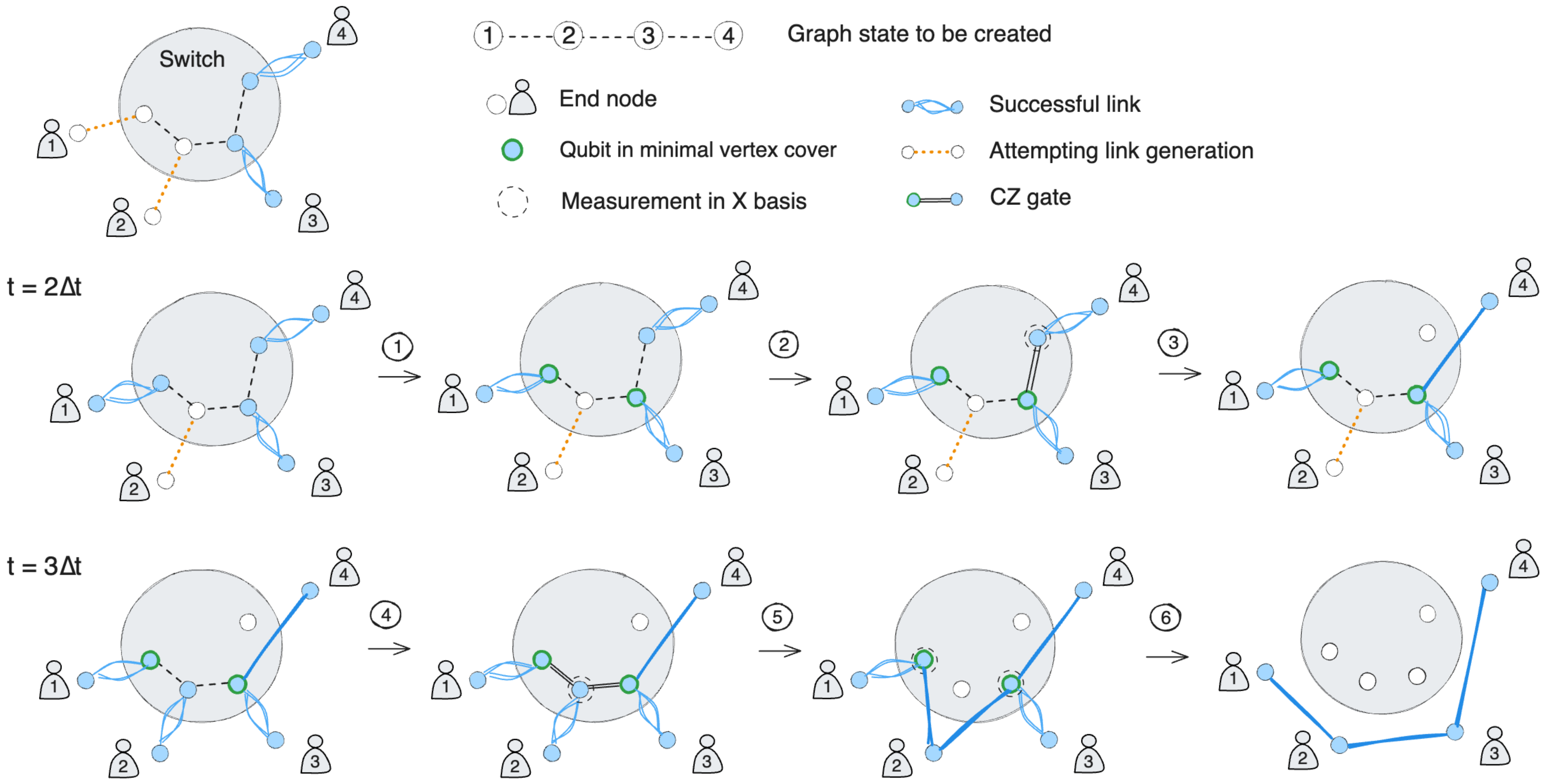}
    \caption{The task is to generate a 4-path graph state among the end nodes. In this example, nodes $3$ and $4$ create an entangled link on their first attempt, while nodes $1$ and $2$ succeed after two and three rounds, respectively. (1) The switch node waits until a MVC appears, which in this case is $\mathrm{MVC}=\set{1,3}$ at $t=2\Delta t$. (2-3) The switch then fuses the qubit of node $4$ into its neighboring qubit $3$ in the vertex cover. (4-5) Once also node $2$ succeeds and the $\CZ$ gates to neighbors $1$ and $3$ have been performed, the switch measures the qubits in the vertex cover. (6) The target graph state is established. Note that only attempts to create links are assumed to take time, while classical operations (e.g., determining if a MVC is present) the $\CZ$ operations and measurements are treated as instantaneous.}
    \label{fig:sketchMVC}
\end{figure*}

We now extend the concept of piece-wise entanglement distribution to general stabilizer states. We introduce two protocols, the minimal vertex cover (MVC) and the (general) Piecemaker protocol. The MVC protocol we introduce as a stepping stone to the Piecemaker protocol, since the MVC protocol can be interpreted as a subroutine of the Piecemaker protocol.

We will phrase our protocols in terms of \emph{graph states} (to be defined later in Def.~\ref{def:graphstate}), a subset of stabilizer states that are easier to study due to their graphical representation. It is well known that any stabilizer state can be transformed into a graph state using single-qubit Cliffords, and the required gates can be found in polynomial time~\cite{van2004graphical}. It therefore suffices to focus on graph states when studying protocols that distribute general stabilizer states.

Before specializing to graph states, let us make the following observation. 
Any protocol for preparing a target state using an entanglement switch, including the GHZ Piecemaker and Factory protocols, effectively does so by performing a collective measurement on the entangled qubits held by the switch (which can be interpreted as a form of remote state preparation~\cite{bennett2001remote}).
As such, it is important to understand which states can arise after performing measurements on subsets of Bell pairs at the switch. Consider the $n$ Bell pairs (or any subset thereof), and note that $\left(\frac{\ket{00}+\ket{11}}{\sqrt{2}}\right)^{\otimes n} =\frac{1}{\sqrt{2^n}}\sum_{i=0}^{2^n-1}\ket{ii}\equiv\ket{\Psi}$ up to relabeling (where the bipartition is between the switch and the $n$ end users). As such, the transpose trick~\cite{wilde2011classical} applies, i.e.~for any operator $A$ of the appropriate size we have $A\otimes I\ket{\Psi}=I\otimes A^T\ket{\Psi}$. Thus, applying projections at the switch has the same effect as applying those projections at the end users. More specifically, by measuring a set of stabilizer generators of a stabilizer state $\ket{\psi}$ at the switch, the state $\ket{\psi}$ is distributed (up to possible corrections) to the end users.

Let us now specialize to graph states.
\begin{definition}[Graph state] \label{def:graphstate}
  Let $G=(V,E)$ be a simple undirected graph on $|V|=n$ vertices.
  The associated $n$-qubit \emph{graph state} is  
  $
    \ket{G} = \Bigl(\prod_{\lbrace u,v\rbrace\in E} \CZ_{uv}\Bigr)
     \ket{+}^{\otimes n},
  $
  where every qubit $v\in V$ is initialized in
  $\ket{+} = (\ket{0}+ \ket{1})/\sqrt2$
  and a $\CZ_{uv}$ gate is applied for each edge $\lbrace{u,v\rbrace}$.
  A convenient choice of stabilizer generators for the graph state is given by
  \[
    K_v \equiv  X_v \prod_{u\in N(v)} Z_u,\quad  \forall v\in V\ ,
  \]
  where $N(v)$ are the neighbors of vertex $v$.
\end{definition}

The stabilizer $K_v$ acts non-trivially only on $v$ and $N(v)$.  Thus, to measure $K_v$ at the switch, it suffices to perform $\CZ$ gates from the qubits $N(v)$ to $v$, and then measure $v$ in $X$. After this measurement, qubit $v$ is no longer entangled with any end user. This has two effects. First, being able to measure qubit $v$ before all Bell pairs have been established reduces the time over which qubit $v$ decoheres, and thus does not reduce the quality of the final state. Secondly, after measuring $K_v$, any other stabilizer $K_w$ cannot be measured if $K_w$ acts non-trivially on $v$. This occurs exactly when $v$ and $w$ are neighbors in the graph state to be distributed, and thus limits when the different stabilizers can be measured.

\subsection{The MVC protocol}
Our first protocol, which we dub the MVC protocol, aims to perform the measurements $K_v$ as soon as possible, with the constraint that any remaining stabilizer measurement $K_w$ can still be performed. Let us make this concrete with an example of a path graph, see Fig.~\ref{fig:sketchMVC}. After end users 1, 3 and 4 have established entanglement with the switch, measurement $K_4$ can be performed. This is because qubit $4$ does not share any edge with any of the Bell pairs that still need to be established (which in this case is qubit $2$). Measuring out qubit $4$ before Bell pair $2$ gets established reduces the total amount of time qubit $4$ is stored in memory, increasing the quality of the final state (compared to storing the latter in memory until the last moment). After end user $2$ has established entanglement with the switch, the remaining stabilizers $K_1$, $K_2$, $K_3$ can be measured. Note that to measure $K_3$, we do not need to perform a $\CZ$ gate from qubit $4$ to $3$, since such a gate was already applied before measuring $K_4$.

Let us now give a general overview of the MVC protocol. First, a vertex cover of a graph $G$ is a subset of vertices $\mathcal{W} \subseteq V$ such that every edge in $G$ is incident with at least one of the vertices in $\mathcal{W}$. The (set) complement of a vertex cover is, by definition, a subset of vertices such that no edge in $G$ is incident on two vertices in that set. As soon as the Bell pairs that have arrived at the switch form a vertex cover, it is possible to perform $\CZ$ gates and measure out any of the qubits that will arrive in the future. This is because---by construction---any two qubits that still need to arrive will not share an edge.

The earlier the created Bell pairs form a vertex cover, the more storage time can be reduced. Furthermore, any superset of a vertex cover is a vertex cover as well. As such, we are interested in \emph{minimal} vertex covers, since as soon as they are established, Bell pairs can start to be measured out.

\begin{definition}[Minimal vertex cover (MVC)] \label{def:mlvc}
  A vertex cover is \emph{minimal} if none of its strict subsets is a vertex cover.
\end{definition}

Note that the measurement outcomes at the switch are random, and---depending on their outcomes---corrections may need to be applied at the end users. We give a full step-by-step description of the MVC protocol in Protocol~\ref{prot:generalpiecemaker}.
In the following, $\ket{G_t}$ represents the target multipartite entangled state to be shared across the $n$ end nodes, with $G_t$ its underlying graph.
We also define $S$ to be the set of switch qubits that have succeeded generating entanglement with remote node qubits, but which have not yet been measured out.

\begin{protocol}[MVC Protocol]\label{prot:generalpiecemaker}
  \item All end nodes in parallel engage in heralded Bell pair generation.
  \item Once a link succeeds, its corresponding switch qubit gets added to the set of current qubits $S$. The switch checks if $S$ forms a vertex cover of the graph $G_t$. If so, the switch finds an MVC $\mathcal{W}\subseteq S$ (this can be done efficiently by removing elements at random from $S$ until a minimal vertex cover is found). Otherwise, if no MVC is present, the switch waits and repeats this step.
  \item Denote the complement of $\mathcal{W}$ in $V$ by $\mathcal{W}^c$. The switch measures the standard graph state stabilizers $K_v$ for $v\in \mathcal{W}^c\cap S$, taking into account that certain $\CZ$ gates may have been performed already to measure other qubits (see main text). Every time a qubit gets measured, remove it from $S$.
  \item The switch waits for the next successful link to be created, and repeats from Step $3$ until all qubits~$j\in\mathcal{W}^c$ have been measured.
  \item All stabilizers $K_v$ for $v \in \mathcal{W}$ are measured, noting that any $\CZ$ gates involving qubits already measured in earlier steps have been applied beforehand. Finally, the classical outcomes are sent to the end nodes for correction.
\end{protocol}

\subsection{The Piecemaker protocol}
The MVC protocol from the previous section might not work well for some graphs. Take for example a complete graph. The minimal vertex covers of a complete graph are of size $|\mathcal{W}| = |V_t|-1$. In this case, the MVC protocol does not provide any benefit over the Factory protocol. However, the complete-graph state is equivalent to a GHZ state up to single-qubit Clifford operations, revealing that a smarter protocol should be able to do much better than the Factory protocol.
This motivates us to modify the MVC protocol to take into account the fact that certain graph states are equivalent to one another, up to single-qubit Clifford rotations. Two graph states are single-qubit Clifford equivalent if and only if $G$ and $G'$ are \emph{locally equivalent}~\cite{van2004graphical}.

\begin{definition}
[Local complementation] \label{def:lcop}
Let $G=(V,E)$ be a simple undirected graph and let $v\in V$. The \emph{local complement} of $G$ at $v$ is the graph with the same edges and vertices as $G$, but where the adjacency relations between the neighbors of $v$  are inverted. In other words, local complementation removes all edges that are present between the neighbors of $v$, and adds the edges that were not present. Two graphs that are related by a sequence of local complementations are called \emph{locally equivalent}, or \emph{local Clifford (LC) equivalent}.
\end{definition}

\begin{figure}
    \centering

\begin{tikzpicture}

\node[circle, fill=black, draw, scale=0.6] (1) at ({sin(0*360/5)}, {-cos(0*360/5)}){};
\node[circle, fill=black, draw, scale=0.6] (3) at ({sin(2*360/5)}, {-cos(2*360/5)}){};
\node[circle, fill=black, draw, scale=0.6] (4) at ({sin(3*360/5)}, {-cos(3*360/5)}){};
\node[circle, fill=white, draw, scale=1] (0) at (0.0, -0){};
\node[circle, fill=black, draw, scale=0.6] (0) at (0.0, -0){};
\draw[line width = 0.3mm] ({sin(2*360/5)}, {-cos(2*360/5)}) -- ({sin(3*360/5)}, {-cos(3*360/5)});

\draw[line width = 0.3mm] (0,-1) -- (0.8, -1.2);
\draw[line width = 0.3mm] (0,-1) -- (-0.7, -1.3);

\draw[line width = 0.3mm] ({sin(2*360/5)}, {-cos(2*360/5)}) -- (1.1, 0.6);
\draw[line width = 0.3mm] ({sin(2*360/5)}, {-cos(2*360/5)}) -- (1.2, 1.3);
\draw[line width = 0.3mm] ({sin(2*360/5)}, {-cos(2*360/5)}) -- (0.4, 1.6);
\draw[line width = 0.3mm] ({sin(3*360/5)}, {-cos(3*360/5)}) -- (-1.1, 0.4);
\draw[line width = 0.3mm] ({sin(3*360/5)}, {-cos(3*360/5)}) -- (-1.2, 1.1);

\draw[line width = 0.3mm] (0,-1) -- (0, -0);

\draw[line width = 0.3mm] ({sin(2*360/5)}, {-cos(2*360/5)}) -- (0, -0);
\draw[line width = 0.3mm] ({sin(3*360/5)}, {-cos(3*360/5)}) -- (0, -0);

\draw [-stealth](2, 0) -- (2.3,0);



\def\ra{4.2}

\node[circle, fill=black, draw, scale=0.6] (1) at ({sin(0*360/5)+\ra}, {-cos(0*360/5)}){};

\node[circle, fill=black, draw, scale=0.6] (3) at ({sin(2*360/5)+\ra}, {-cos(2*360/5)}){};
\node[circle, fill=black, draw, scale=0.6] (4) at ({sin(3*360/5)+\ra}, {-cos(3*360/5)}){};

\node[circle, fill=white, draw, scale=1] (0) at (0.0+\ra, -0){};
\node[circle, fill=black, draw, scale=0.6] (0) at (0.0+\ra, -0){};

\draw[line width = 0.3mm] ({sin(2*360/5)+\ra}, {-cos(2*360/5)}) -- ({0+\ra}, {-1});
\draw[line width = 0.3mm] ({sin(3*360/5)+\ra}, {-cos(3*360/5)}) -- ({0+\ra}, {-1});

\draw[line width = 0.3mm] (0+\ra,-1) -- (0.8+\ra, -1.2);
\draw[line width = 0.3mm] (0+\ra,-1) -- (-0.7+\ra, -1.3);

\draw[line width = 0.3mm] ({sin(2*360/5)+\ra}, {-cos(2*360/5)}) -- (1.1+\ra, 0.6);
\draw[line width = 0.3mm] ({sin(2*360/5)+\ra}, {-cos(2*360/5)}) -- (1.2+\ra, 1.3);
\draw[line width = 0.3mm] ({sin(2*360/5)+\ra}, {-cos(2*360/5)}) -- (0.4+\ra, 1.6);
\draw[line width = 0.3mm] ({sin(3*360/5)+\ra}, {-cos(3*360/5)}) -- (-1.1+\ra, 0.4);
\draw[line width = 0.3mm] ({sin(3*360/5)+\ra}, {-cos(3*360/5)}) -- (-1.2+\ra, 1.1);

\draw[line width = 0.3mm] (0+\ra,-1) -- (0+\ra, -0);

\draw[line width = 0.3mm] ({sin(2*360/5)+\ra}, {-cos(2*360/5)}) -- (0+\ra, -0);
\draw[line width = 0.3mm] ({sin(3*360/5)+\ra}, {-cos(3*360/5)}) -- (0+\ra, -0);

\end{tikzpicture}
    \vspace{-0.8cm}
    \caption{Example of a local complementation on the circled vertex $v$. The adjacency relations of the neighbors of $v$ are reversed i.e., if there was an (no) edge between two neighbors of $v$, it is removed (added). Example taken from~\cite{goodenough2023near}.}
    \label{fig:lcexample}
\end{figure}
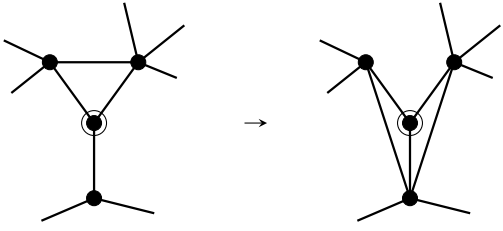

We show an example of a local complementation in Fig.~\ref{fig:lcexample}.

The core idea is as follows. Assume the switch wants to distribute a target state $\ket{G_t}$ to the end users. We are free to distribute any locally equivalent graph state $\ket{G'}$ instead, since end users can perform single-qubit operations to transform $\ket{G'}$ into $\ket{G_t}$. The graph state $\ket{G'}$ is chosen such that the set $S$ of Bell pairs at the switch forms a (minimal) vertex cover of $G'$. As such, the MVC protocol can be performed for the state $\ket{G'}$, after which the end users transform $\ket{G'}$ back into $\ket{G_t}$.

We now provide two examples. The first one shows that the general Piecemaker protocol is a proper generalization of the GHZ Piecemaker protocol. The second illustrates the general Piecemaker for a more complex state, namely a four-qubit path graph state.

For the first example, note that the GHZ state on qubits $V$ is single-qubit-Clifford equivalent to all star graph states on $V$~\cite{van2004graphical}. As soon as the first Bell pair $v$ arrives, the switch proceeds to run the MVC protocol for the star graph whose central vertex is $v$. This ensures that any Bell pair $w$ that gets created afterwards can be immediately measured out, exactly as in the GHZ Piecemaker protocol. After the star-graph state has been distributed to the end users, single-qubit Cliffords can be performed to turn the state into a GHZ state. Moreover, as the complete graph is also LC equivalent to all star graphs, the general Piecemaker is able to distribute the complete-graph state just as efficiently as the GHZ state.
Before providing the example of the path graph state, we introduce some useful notation. 

\begin{definition}
Let $G$ be a graph on $V$. A subset of vertices $\mathcal{U}\subseteq V$ is called a \emph{local cover} of $G$ if $\mathcal{U}$ is a vertex cover of some graph $G'$ that is locally equivalent to $G$. Any superset of a local cover is a local cover; the local covers that are minimal with respect to set inclusion are called \emph{minimal local covers} (MLC). Finally, the set of minimal local covers of a graph $G$ is denoted by $\mathcal{C}(G)$.
\end{definition}

\begin{figure*}
  \centering
  \includegraphics[width=.8\textwidth]{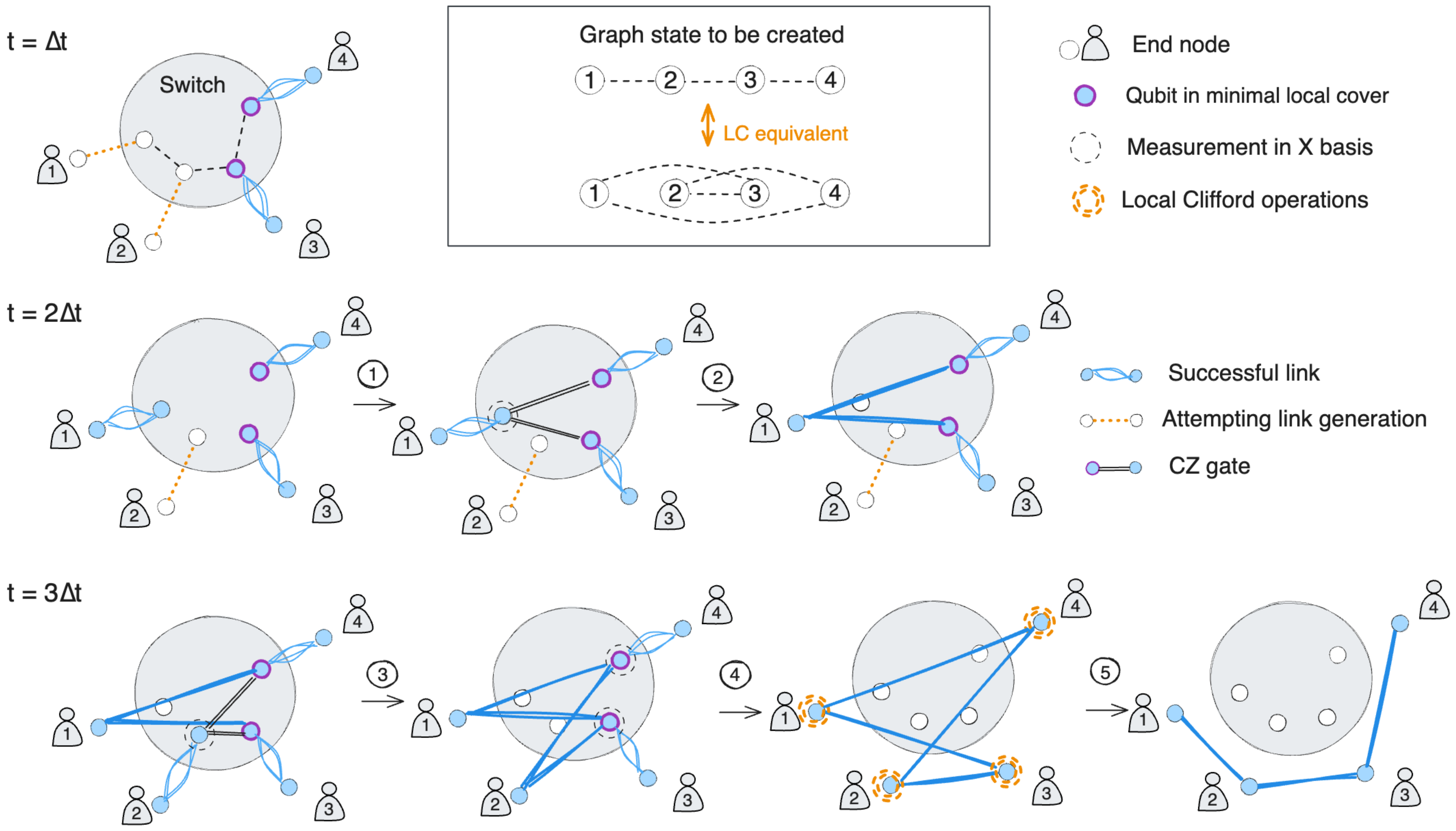}
  \caption{Example of the general Piecemaker protocol utilizing LC equivalence of graph states. Similar to the scnenario in Figure~\ref{fig:sketchMVC}, the task is to generate a path graph state among four end nodes. The set $\set{3,4}$ forms a MLC with the associated LC-equivalent graph $G'(V,E')$, where $E' = \set{\set{1,3}, \set{1,4}, \set{2,3}, \set{2,4}}$ (boxed). (1-2) While in the previous scenario qubit $1$ was part of the MVC, here it can be immediately fused into the MLC qubits. (3) Once the last end node $2$ creates its link at $t=3\Delta t$ ms, the qubits in the MLC are released forming $G'$. (4-5) Finally, the switch recovers the target graph by applying the necessary Clifford operations.}
  \label{fig:sketchLC}
\end{figure*}

Consider the case of a path graph on four qubits in Fig.~\ref{fig:sketchLC}. At $t=\Delta t$ the Bell pairs $3$ and $4$ are present at the switch. The corresponding vertices form an MLC (but not a vertex cover) of the path graph. This is because the cycle graph on four vertices is LC equivalent to the path graph (see~\cite{hein2006entanglement} and the inset of Fig.~\ref{fig:sketchLC}). This cycle graph has $\lbrace{3, 4\rbrace}$ as a vertex cover, and as such we can run the MVC protocol for the cycle graph. In time step $t=2\Delta t$ qubit $1$ can be measured out (i.e., the stabilizer $K_1$ of the cycle graph can be measured), and at $t=3\Delta t$ qubit $2$ is measured out. Afterwards, $\CZ$ gates are applied to perform the remaining stabilizer measurements. Finally, the end users perform the required single-qubit operations to transform the state into the desired path graph state.

We give a general description of the general Piecemaker protocol in Protocol~\ref{prot:generalpiecemakerLC}. As before, $\ket{G_t}$ is the target state, and $S$ represents the set of remotely-entangled but not yet measured switch qubits.

\begin{protocol}[The general Piecemaker protocol] \label{prot:generalpiecemakerLC}
\item Compute $\mathcal{C}(G_t)$.
  \item All end nodes in parallel engage in heralded Bell pair generation. 
  \item Once an end node succeeds, it gets added to the set of current qubits $S$. The switch checks if $S$ forms a \emph{local} cover of the target graph $G_t$. If so, the switch finds an MLC $\mathcal{V}\subseteq S$, and fixes a graph $G'$ that is LC equivalent to $G_t$ and furthermore has $\mathcal{V}$ as a vertex cover. Checking whether $S$ forms a local cover and finding a minimal local cover in $S$ can be done efficiently by randomly removing elements from $S$ until a minimal local cover $\mathcal{V}\in \mathcal{C}(G_t)$ is found.
  Otherwise, if no MLC is present, the switch waits and repeats this step.
  \item Denote the complement of $\mathcal{V}$ in $V$ by $\mathcal{V}^c$. Let $K_v'$ be the standard generators for the graph state $\ket{G'}$, see Def.~\ref{def:graphstate}. The switch performs measurements $K_v'$ for $v\in \mathcal{V}^c\cap S$, taking into account that certain $\CZ$ gates may have been performed already to measure other qubits (see main text). Every time a qubit $v$ gets measured, remove it from $S$. 
  \item The switch waits for the next successful link to be created, and repeats from Step $3$ until all qubits~$j\in\mathcal{V}^c$ have been measured.
  \item All stabilizers $K_v'$ for $v \in \mathcal{W}$ are measured, noting that any $\CZ$ gates involving qubits already measured in earlier steps have been applied beforehand.
  \item End nodes perform the necessary local Clifford operations to transform $\ket{G'}$ into $\ket{G_t}$~\cite{van2004efficient}.
\end{protocol}

Note that the operations in the last step of the protocol can be found from the results from~\cite{van2004efficient}. Furthermore, it is thus possible to adjust the protocol such that the end nodes do not require single-qubit corrections at the end. This is possible because these rotations can be absorbed into the stabilizer measurements $K_v'$ at the switch. We do not consider this modification in the remainder of this paper for simplicity.

Finally, we note that it is not trivial to generate the full set of minimal local covers $\mathcal{C}(G_t)$. In Section~\ref{sec:genUG} we provide an (inefficient) algorithm (available at~\cite{Prielinger2025Data}) that suffices for the system sizes under consideration, especially since the minimal local covers can be generated before running the protocol. Our algorithm uses the algorithm from~\cite{sharma2025minimizing}, which can minimize the (weighted) edges in graph states up to local complementation. We leave open whether there are faster algorithms, either approximate or exact.

\section{Baseline Protocol and numerical evaluation} \label{sec:numerical}
We compare the performance of our protocols to the Factory protocol introduced in~\cite{avis2023analysis}. Here, the switch serves as the Factory node. It contains $2n$ memories, $n$ engaged in link generation, the other $n$ are auxiliary qubits to create the target state $\ket{G_t}$ upon generation of all links. Throughout this section we will sometimes refer to the (GHZ) Piecemaker Protocol simply as "(GHZ) Piecemaker", and to the Factory protocol as "Factory", for short.

{\begin{protocol}[Factory protocol]
  \item All end nodes in parallel engage in heralded Bell pair generation.
  \item The switch node waits until all end nodes have created a link. 
  \item Once all end nodes share a link, the switch creates the target state $\ket{G_t}$ on the auxiliary qubits.
  \item The factory node performs $n$ Bell state measurements, each between one qubit that holds part of the target state, and one qubit that holds one half of a Bell pair in memory $m_i$, where $i\in \set{1,2,\dots,n}$.
  \item The factory node sends the measurement outcomes to the corresponding end nodes, where the associated Pauli $X$ and $Z$ corrections are applied.
\end{protocol}}

In this section we perform an extensive numerical evaluation of the protocols we have introduced and compare them to the Factory protocol. To this end we use the quantum network simulator \textit{QuantumSavory.jl}~\cite{quantumsavoryjl} to account for the non-trivial effect of the random timing of when entangled links are successful on the noise in the distributed state. The goal of our numerical evaluation is to quantify how the time induced by waiting for links degrades states across different protocols.
It will help us understand in which parameter regimes the use of the Piecemaker protocols is most beneficial over the Factory scheme. 

In our simulation, we set $\Delta t = 1$ ms and use a parameter value pair $(p_{\mathrm{link}}, p_{\mathrm{depol}}) \in \mathcal{P}$, where $\mathcal{P}$ is a $20 \times 20$ logarithmically spaced grid over $[10^{-3}, 1]^2$. For each experiment, we report on a set of selected parameter-value pairs $P \subseteq \mathcal{P}$.
 Upon completion of a protocol, the end nodes share a mixed state $\rho^{\mathrm{d,l}}_{\texttt{prot}}$, generated either by the Factory ($\rho_{\texttt{factory}}^{\mathrm{d,l}}$) or the Piecemaker ($\rho_{\texttt{pm}}^{\mathrm{d,l}}$) protocol, where the superscripts $(\mathrm{d,l})$ denote a parameter value pair $(p_{\mathrm{depol}},p_{\mathrm{link}})$. We denote the fidelity to the ideal target state $\ket{G_t}$ as
\begin{equation} \label{eq:fidelity}
  F^{\mathrm{d,l}}_{\texttt{prot}} = \bra{G_t}\rho^{\mathrm{d,l}}_{\texttt{prot}}\ket{G_t} \in [0,1].
\end{equation}
We obtain an estimate of $F^{\mathrm{d,l}}_{\texttt{prot}}$, denoted by $\tilde{F}^{\mathrm{d,l}}_{\mathrm{prot}}$, by taking a mean over at least $N= 10^4$ Monte Carlo trials.

To compare the Piecemaker variants with the baseline Factory protocol we report the approximated \emph{difference in fidelities}
\begin{equation} \label{eq:diffdl}
  \Delta \tilde{F}^{\mathrm{d,l}} := \tilde{F}^{\mathrm{d,l}}_{\texttt{pm}} - \tilde{F}^{\mathrm{d,l}}_{\texttt{factory}},
\end{equation}
and the \emph{relative reduction of infidelity}
\begin{equation} \label{eq:errdl}
  \Delta \tilde{\epsilon}^{\mathrm{d,l}} := \frac{\tilde{F}^{\mathrm{d,l}}_{\texttt{pm}} - \tilde{F}^{\mathrm{d,l}}_{\texttt{factory}}}{1 - \tilde{F}^{\mathrm{d,l}}_{\texttt{factory}}},
\end{equation}
i.e., the fraction by which the infidelity, which can be interpreted as the probability that the delivered state is not the target state, decreases when switching from Factory to Piecemaker.

We denote overall fidelity average and average improvements using
\begin{equation}
    \overline{F}_\texttt{prot} = \frac{1}{|P|}\sum_{(\mathrm{d,l})\in P} \tilde{F}_\texttt{prot}^{\mathrm{d,l}}, 
\end{equation}
\begin{equation}
\Delta \overline{F} = \frac{1}{|P|}\sum_{(\mathrm{d,l})\in P}\Delta \tilde{F}^{\mathrm{d,l}}  \quad\text{and}\quad \Delta \overline{\epsilon} = \frac{1}{|P|}\sum_{(\mathrm{d,l})\in P}\Delta\tilde{\epsilon}^{\mathrm{d,l}},
\end{equation}
with overall largest improvements as
\begin{equation}
    \Delta F^* = \max_{\mathrm{d,l}}\Delta \tilde{F}^{\mathrm{d,l}} \ \mathrm{and}\ \Delta \epsilon^* = \max_{\mathrm{d,l}}\Delta\tilde{\epsilon}^{\mathrm{d,l}}.
\end{equation}

We compare our results to the findings reported in~\cite{avis2023analysis}. In particular, we compare the performance of
\begin{enumerate}
  \item the GHZ Piecemaker protocol to the Factory protocol for system sizes up to $n=50$ end nodes;
  \item the MVC and general Piecemaker protocols to the Factory protocol for path and grid graphs of up to $n=50$, and a set of example graph states reported in~\cite{adcock2020mapping} for system sizes up to $n=8$ end nodes, respectively.
\end{enumerate}
Note that we do \textit{not} analyze rates, i.e., number of generated states per unit time, as distribution rates are identical to those reported for  the Factory protocol in\cite{avis2023analysis}.

\subsection{GHZ Piecemaker}
\label{sec:ghzresults}
In the GHZ Piecemaker we perform piece-wise fusion of the incoming Bell pairs to create a shared $n$-GHZ state among $n$ end nodes. We first investigate the overall behavior of the setup when all end nodes are located at the same distance from the switch (\emph{homogeneous setup}). Subsequently, we will test the protocol over varying link lengths (\emph{inhomogeneous setup}).

\begin{figure}[t]
\centering
\includegraphics[width=\linewidth]{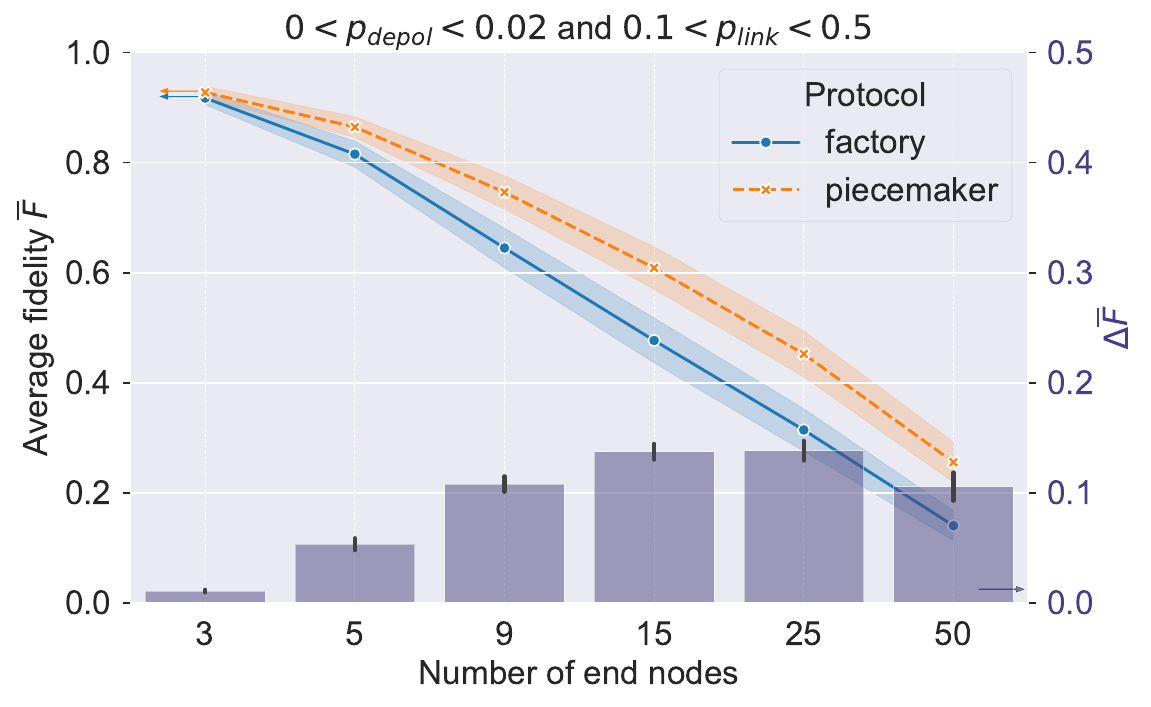}
\caption{Overview of the performance of the GHZ Piecemaker protocol compared to the Factory protocol for system sizes up to $n=50$ qubits. Lines indicate the average fidelity $\overline{F}_\texttt{prot}$ in parameter ranges $0.1 < p_{\mathrm{link}} < 0.5$ and $p_{\mathrm{depol}} < 0.02$. The bars (purple) indicate the average difference $\Delta \overline{F}$ between the two protocols within these parameter intervals.}
\label{fig:ghzresults_overview}
\end{figure}

\begin{figure*}[ht]
\centering
\includegraphics[width=.8\linewidth]{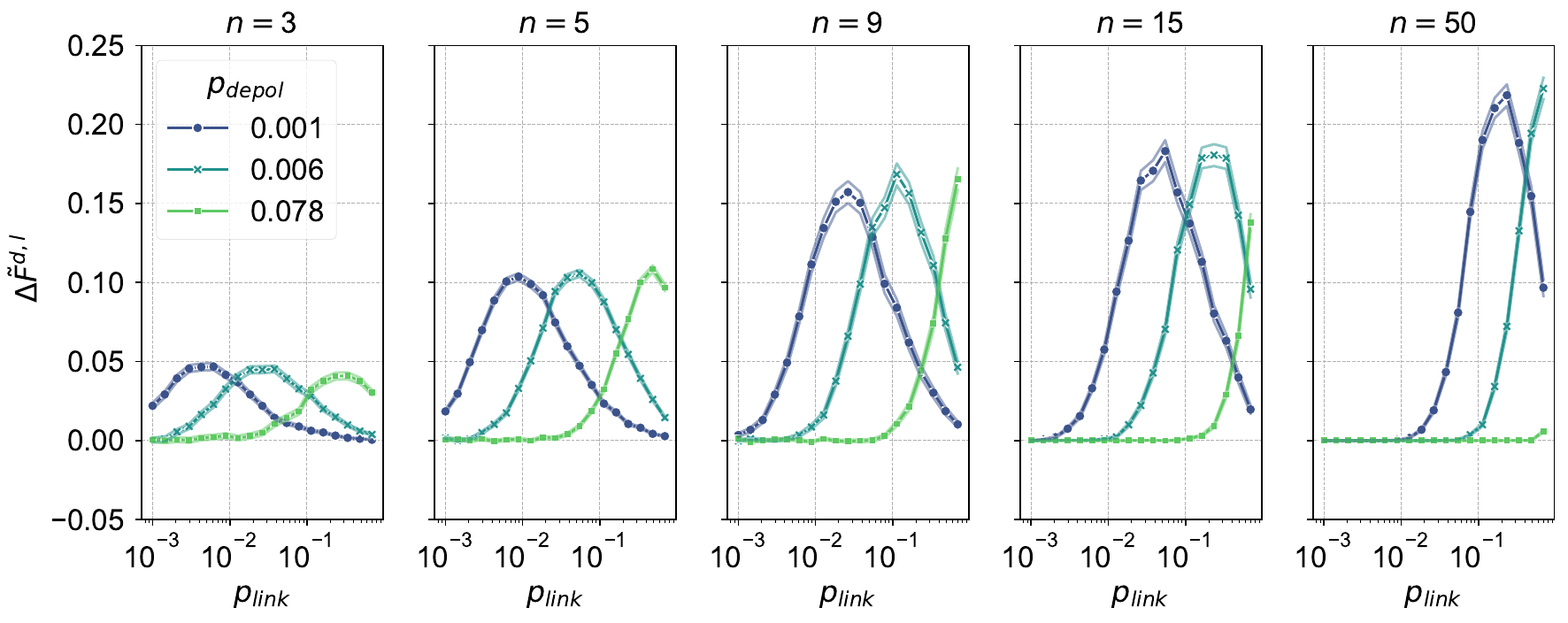}
\caption{Fidelity difference $\Delta \tilde{F}^{\mathrm{d,l}}$~(\ref{eq:diffdl}) between the GHZ Piecemaker protocol and the Factory protocol for system sizes up to $n=50$ qubits. Error envelopes indicate the standard error of the mean of $N=10^5$ runs for $n=\set{3,5}$ and $N=10^4$ runs for $n>5$.
}
\label{fig:ghzresults_difference}
\end{figure*}

\begin{figure*}[ht]
\centering
\includegraphics[width=.8\linewidth]{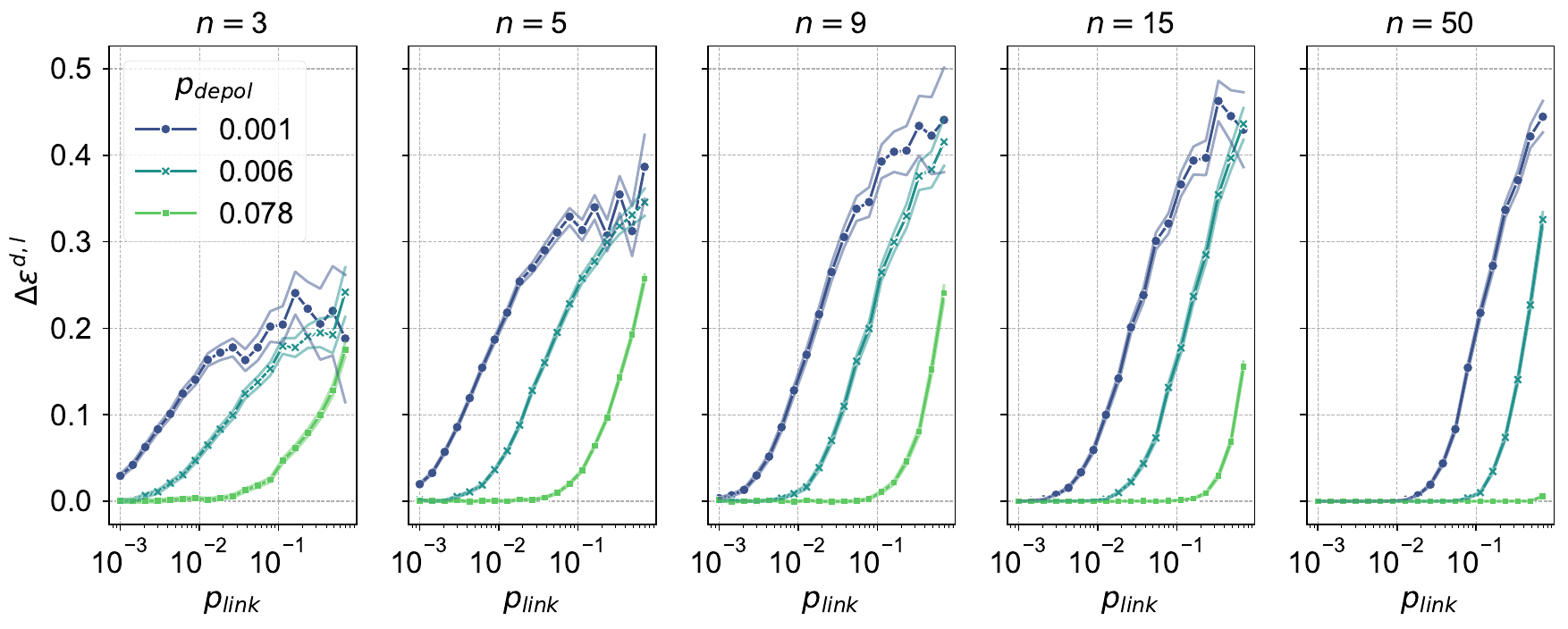}
\caption{Relative decrease in infidelity $\Delta \tilde{\epsilon}^{\mathrm{d,l}}$~(\ref{eq:errdl}) for system sizes up to $n=50$ qubits. Error envelopes indicate the standard error of the mean of $N=10^5$ runs for $n=\set{3,5}$ and $N=10^4$ runs for system sizes $n>5$.}
\label{fig:ghzresults_infidelity}
\end{figure*}

\subsubsection{Homogeneous setup}
We first investigate the performance of the GHZ Piecemaker protocol for system sizes up to $n=50$ qubits. An overview of the average fidelities $\overline{F}_\texttt{prot}$ for Piecemaker and Factory are shown in Figure~\ref{fig:ghzresults_overview} for a selected $P$ of link success probabilities and memory depolarization rates, $0.1 < p_{\mathrm{link}} < 0.5$ and $0 < p_{\mathrm{depol}} < 0.02$. 
In the displayed parameter regimes the largest fidelity improvements appear for system sizes $n=\set{15,25}$ with $\Delta \overline{F} \approx  0.13$.

When we turn to individual parameter settings in Figures~\ref{fig:ghzresults_difference} and~\ref{fig:ghzresults_infidelity}, the largest performance improvement
is observed for $n=50$ and $p_{\mathrm{depol}}\le 0.006$, achieving ${\Delta F^* \approx 0.22}$ and $\Delta \epsilon^*\approx 0.45$.
We observe in Figure \ref{fig:ghzresults_difference} that the fidelity difference $\Delta \tilde{F}^\mathrm{d,l}$ for $p_\mathrm{depol}\le 0.006$ increases with system size. The Piecemaker protocol is able to prevent an increasingly large fraction of errors in this small error probability regime by releasing memories early.

Furthermore, Figure \ref{fig:ghzresults_difference} shows the fidelity difference $\Delta \tilde{F}^{\mathrm{d}, \mathrm{l}}$ forms a bell-shaped curve over increasing link success probabilities: at low link success probabilities, depolarization dominates due to long generation times, and Piecemaker is not able to surpass the performance of Factory. As link-success probability rises, generation time and therefore overall errors decrease, and Piecemaker gains a clear advantage since the positive effect of earlier qubit measurements becomes prevalent. Once link-success probability approaches 1 and depolarization becomes negligible, the gap between the protocols closes again.

The relative decrease in infidelity $\Delta \tilde{\epsilon}^{\mathrm{d,l}}$, reveals in Figure~\ref{fig:ghzresults_infidelity}  a subtler benefit: the Piecemaker achieves a sizable relative infidelity reduction even across the high link success probability regime. For depolarization rates $p_\mathrm{depol}=0.006$, for example, the Piecemaker achieves a relative reduction in infidelity of more than 40\% in systems with $n = 9$ and $n=15$ end nodes.

\begin{figure*}
    \centering
    \includegraphics[width=.9\linewidth]{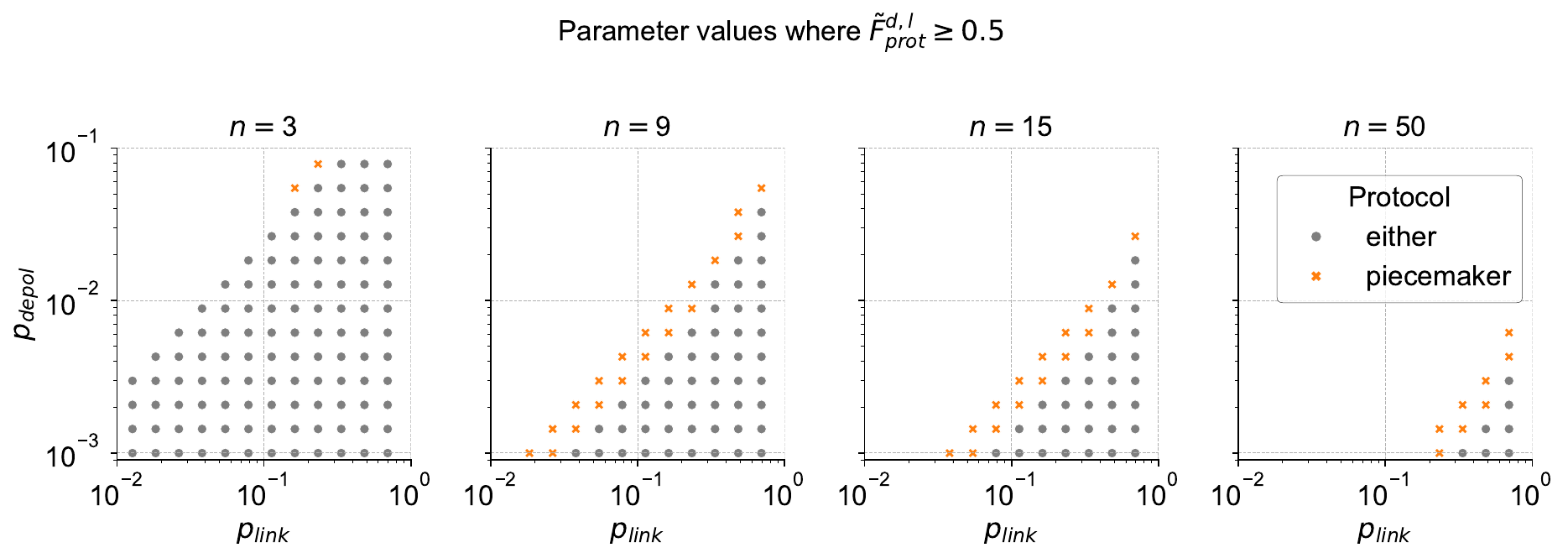}
    \caption{Parameter values where the final graph state has a fidelity of $\tilde{F}_{\texttt{prot}}^{\mathrm{d,l}}\ge \frac{1}{2}$. Empty spaces indicate parameter regimes where neither protocol could reach the threshold.}
    \label{fig:ghzabove0.5}
\end{figure*}

When it comes to GHZ states, a fidelity of above $\frac{1}{2}$ guarantees genuine multipartite entanglement~\cite{guhne2009entanglement}. That is, one can construct an entanglement witness able to prove experimentally that a degraded GHZ state is still entangled when the fidelity to the ideal state is $\frac{1}{2}$ or higher. Figure \ref{fig:ghzabove0.5} shows in which parameter regime the protocols can achieve this threshold. Naturally, the larger the system size the smaller the scope of parameter values with which this threshold can be reached. Nonetheless, across all system sizes the Piecemaker meets the fidelity target over a noticeably broader range of parameters than the baseline. Remarkably, for a depolarization probability of about 0.005, the minimal required link success probability is about 0.1 while for the Factory scheme this parameter should be at least 0.2. We report on further results for a fidelity threshold of 0.9, which are included in the accompanying data repository~\cite{Prielinger2025Data}. For a network of nine remote nodes, for example, the minimal required link success probability to achieve a threshold fidelity of 0.9 can fall from 0.23 to 0.16 when the depolarization probability is fixed at 0.001. Conversely, if the link success probability is held at 0.7, the minimal required depolarization probability can rise from 0.004 to 0.009. Even though these differences may not appear large, they can be decisive for the the possibility of realizing such a protocol considering current hardware capabilities~\cite{pont2024high}.

\subsubsection{Heterogeneous setup}
Previous studies have observed that the performance of quantum-repeater chains is degraded when the losses are distributed in an inhomogeneous fashion across links~\cite{avis2023asymmetric}.
We expect similar behavior in the case of entanglement switches with inhomogeneous links, 
where entangled link inter-arrival times usually have a higher variance than in the homogeneous link case.
To this end, we define a physically motivated model, inspired by \cite{avis2023asymmetric}, for redistributing losses across the different links of the star network such that the total amount of loss (i.e., the product of the loss parameters) remains constant.
A simple model for relating link loss with the length $L_i$ of a fiber optic channel connecting end node $i$ to the switch can be written as
$$
p^i_{\mathrm{link}} \approx 10^{-\gamma \cdot L_i / 10},
$$
where $p^i_{\mathrm{link}}$ is the link success probability and $\gamma = 0.2\ \mathrm{dB/km}$ is the fiber attenuation coefficient~\cite{rozpkedek2019near, coopmans2021netsquid}.
This expression corresponds to the direct transmission of the photons through the optical fibers, and on a qualitative level captures the exponential scaling common to many heralded entanglement generation schemes, including the single-click~\cite{cabrillo1999creation} and double-click~\cite{barrett2005efficient} schemes.
We set the distance from the switch to end node $i\in\set{1,\ldots ,n}$ as $L_i = 25 \text{ km} + (i-3)\Delta L$, where $\Delta L \in \set{1,\dots,10}$ km. 
Note that $\sum_i L_i$ is invariant across $\Delta L$, i.e., the total amount of fiber remains the same in all experiments. The $\Delta L$ parameter thus enables tuning of link inhomogeneity while keeping aggregate loss in the network constant. We assume that---as in the homogeneous-link case---an entanglement generation attempt consumes $\Delta t=1$ ms regardless of link length (even if in principle shorter links could fit more attempts in each period). This assumption is compatible with the longer link length we consider: it takes approximately $0.225$ ms for light to traverse $45$ km in optical fiber. This model of link generation helps isolate the effect of inhomogeneous losses.

\begin{figure}
    \centering
    \includegraphics[width=\linewidth]{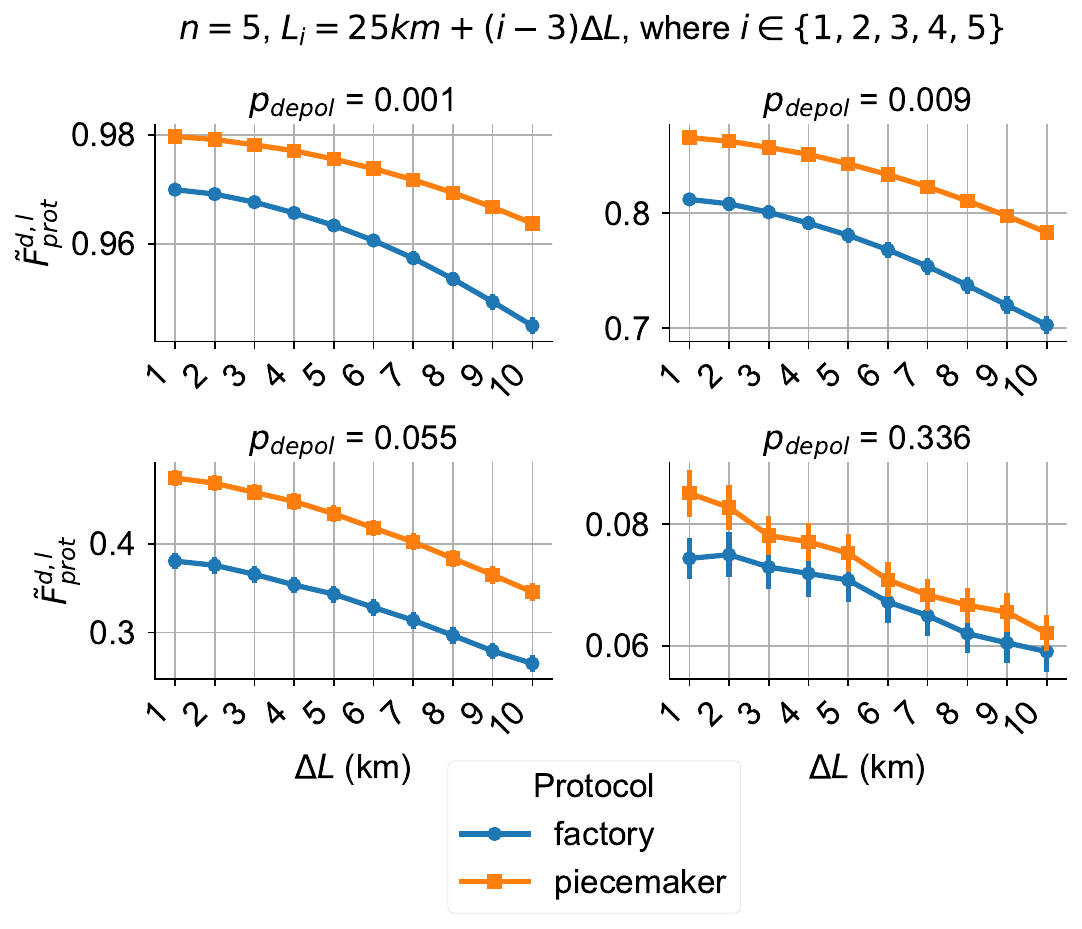}
    \caption{Final fidelities $\tilde{F}^{\mathrm{d,l}}_\texttt{prot}$ of GHZ Piecemaker and Factory protocol over varying link length differences $\Delta L$. Error bars indicate the standard error of $N=10^4$ protocol executions.
    }
    \label{fig:5GHZheterogen}
\end{figure}

Figure~\ref{fig:5GHZheterogen} shows the final fidelities $\tilde{F}^\mathrm{d,l}_\texttt{prot}$ over varying link length differences $\Delta L$.
Overall, we observe from these results  that when memory noise is low and the final fidelity exceeds $\frac{1}{2}$, Piecemaker is less sensitive to inhomogeneity than Factory. E.g., take the case in Figure~\ref{fig:5GHZheterogen} when $p_{\mathrm{depol}} = 0.001$: fidelity decreases by $1.6\%$ under Piecemaker over the growing distance $\Delta L$, while the decrease under Factory is $2.5\%$.  This difference diminishes when $p_\mathrm{depol}\ge 0.05$, and fidelities are below $\frac{1}{2}$.

\subsection{General Piecemaker}
The general Piecemaker protocol allows us to explore arbitrary graph states. We compare its performance against the Factory protocol for different target graph states.

\subsubsection{Complete graph} As noted before, a complete graph on $n$ vertices is LC equivalent to any star graph on $n$ vertices. A complete graph is thus also LC equivalent to a GHZ state~\cite{van2004graphical}. Given a complete graph as target state, the MVC variant of Piecemaker exhibits equivalent performance to that of Factory, since we have to wait until all but one link have succeeded. The general Piecemaker on the other hand leads to performance gains equivalent to those of GHZ Piecemaker; and it therefore achieves the same improvement in fidelity, $\Delta {F}^* = 0.22$,  and a relative decrease in infidelity of up to $\Delta \epsilon^* = 0.45$.

\subsubsection{Grid and path graph} 
\begin{figure*}[ht]
\centering
\includegraphics[width=.7\linewidth]{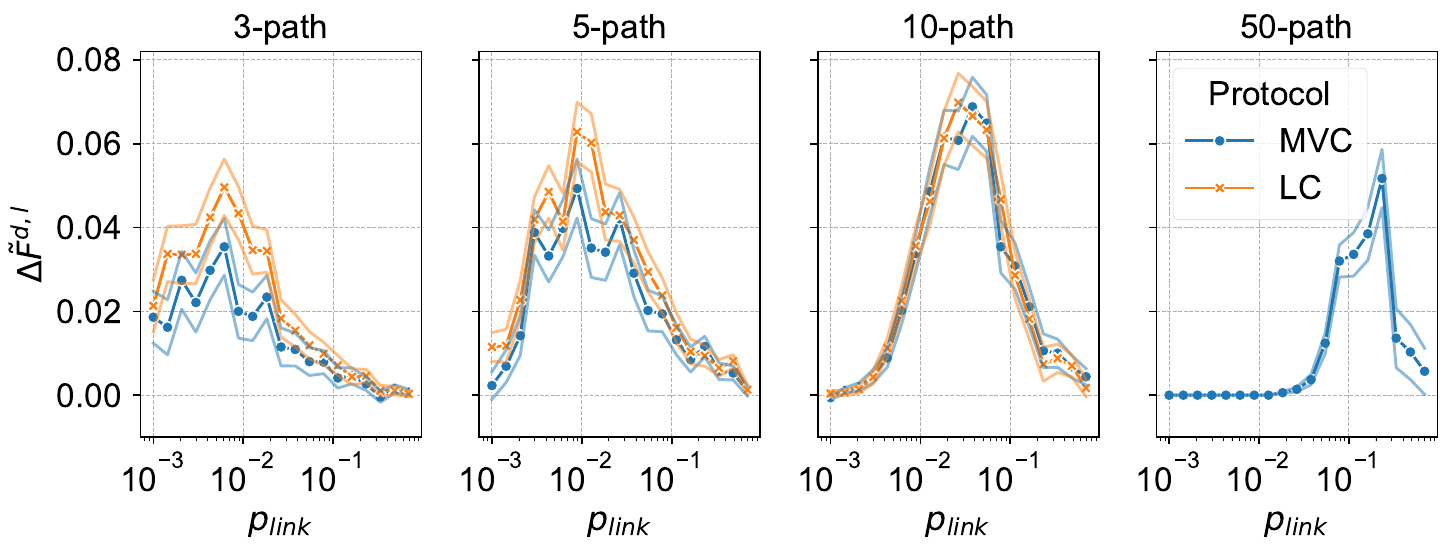}\\
\includegraphics[width=.7\linewidth]{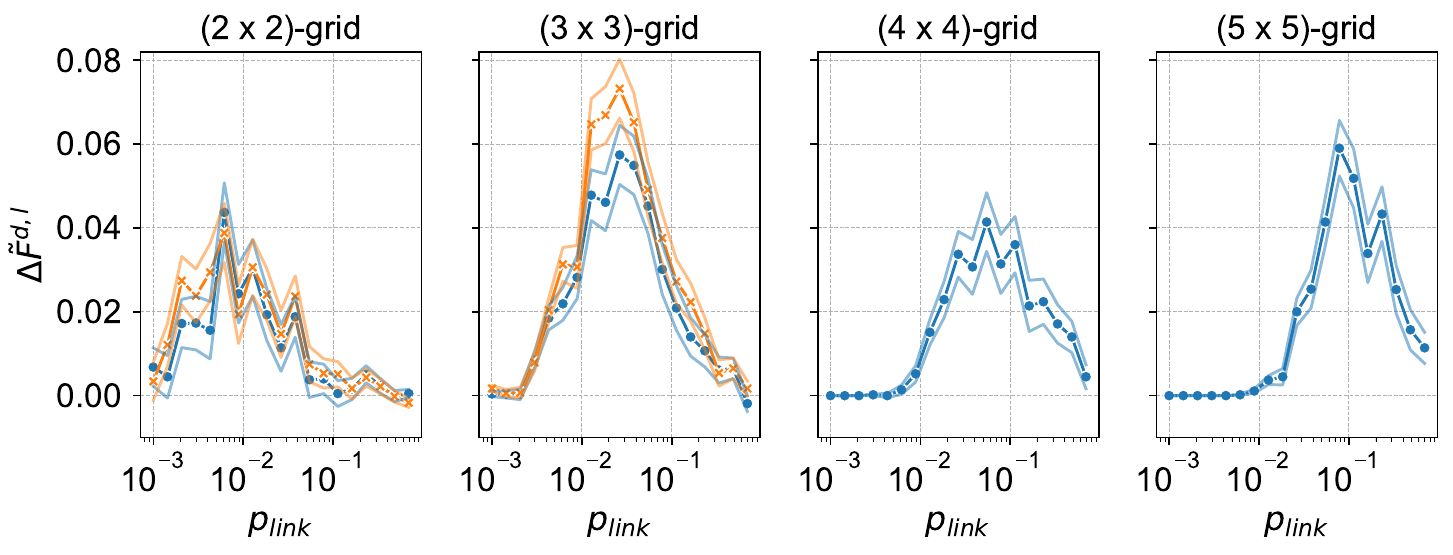}
\caption{Fidelity difference $\Delta \tilde{F}^{\mathrm{d,l}}$ between the general Piecemaker protocols and the Factory protocol, where the depolarizaton probability is $p_{\mathrm{depol}} = 0.001$. Note that we evaluate system sizes up to $n=10$ when using the general Piecemaker (LC, orange line), due to the computational complexity of finding MLCs \cite{sharma2025minimizing}. The error envelopes indicate the standard error of the mean over $N=10^4$ simulation runs.}
\label{fig:gridpath_fidelity_difference}
\end{figure*}

A path graph is a simple graph consisting of vertices arranged in a single line, where each vertex (except the two endpoints) is connected to exactly two neighbors. A grid graph, also known as two-dimensional lattice graph, is the graph Cartesian product of two path graphs forming a lattice of vertices with edges between horizontally and vertically adjacent vertices\cite{acharya1981index}. In both grid and path graphs MVCs grow in general with the size of the graph. For this reason the expected improvement over Factory is lower. Figure~\ref{fig:gridpath_fidelity_difference} shows the average fidelity difference $\Delta \tilde{F}^{\mathrm{d,l}}$ between protocols for path and grid graphs of system sizes up to $n=50$ end nodes and $n=25$ end nodes, respectively. The depolarizing rate is kept constant at $p_{\mathrm{depol}} = 0.001$. Final fidelities produced by the  MVC variant are equal or slightly below the final fidelities generated by the general Piecemaker protocol. 
However they are not negligible with $\Delta F^*$ as large as $=0.074$, and $\Delta \epsilon^* =19\%$.

\subsubsection{Other graph examples}
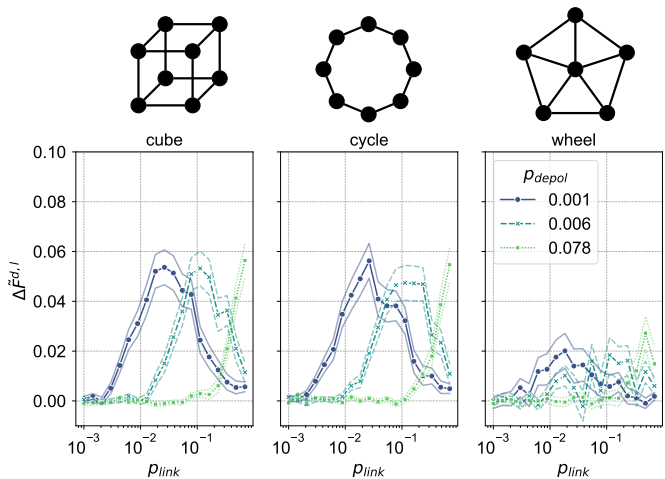
\begin{figure}
\centering
\begin{tikzpicture}[scale=0.6]

\node[circle, fill=none, draw=none, scale=0.6] (borderleft) at (-1, 1.0)      {};

\node[circle, fill=none, draw=none, scale=0.6] (borderright) at (13, 1.0)      {};

\def\shiftcubeX{2.2}    
\def\shiftcubeY{-0.2}    

\def\shiftcycleX{6.7} 
\def\shiftcycleY{0}   

\def\shiftwheelX{11.28} 
\def\shiftwheelY{0}    

\def\dz{-0.6}           

\node[circle, fill=black, draw, scale=0.6] (b1) at (\shiftcubeX+0, \shiftcubeY+0)      {};
\node[circle, fill=black, draw, scale=0.6] (b2) at (\shiftcubeX+1.2, \shiftcubeY+0)    {};
\node[circle, fill=black, draw, scale=0.6] (b3) at (\shiftcubeX+1.2, \shiftcubeY+1.2)  {};
\node[circle, fill=black, draw, scale=0.6] (b4) at (\shiftcubeX+0, \shiftcubeY+1.2)    {};

\node[circle, fill=black, draw, scale=0.6] (t1) at (\shiftcubeX+\dz, \shiftcubeY+\dz)  {};
\node[circle, fill=black, draw, scale=0.6] (t2) at (\shiftcubeX+1.2+\dz, \shiftcubeY+\dz) {};
\node[circle, fill=black, draw, scale=0.6] (t3) at (\shiftcubeX+1.2+\dz, \shiftcubeY+1.2+\dz) {};
\node[circle, fill=black, draw, scale=0.6] (t4) at (\shiftcubeX+\dz, \shiftcubeY+1.2+\dz)  {};

\draw[line width=0.3mm] (b1)--(b2)--(b3)--(b4)--(b1);
\draw[line width=0.3mm] (t1)--(t2)--(t3)--(t4)--(t1);
\draw[line width=0.3mm] (b1)--(t1);
\draw[line width=0.3mm] (b2)--(t2);
\draw[line width=0.3mm] (b3)--(t3);
\draw[line width=0.3mm] (b4)--(t4);

\def\r{1.}              

\foreach \i in {0,...,7}{
  \node[circle, fill=black, draw, scale=0.6] (c\i)
    at ({\shiftcycleX+\r*sin(45*\i)}, {\shiftcycleY+\r*cos(45*\i)}) {};
}

\foreach \i in {0,...,6}{
  \pgfmathtruncatemacro{\j}{\i+1}
  \draw[line width=0.3mm] (c\i)--(c\j);
}
\draw[line width=0.3mm] (c7)--(c0);

\def\rc{1.2}            

\foreach \i in {0,...,4}{
  \node[circle, fill=black, draw, scale=0.6] (w\i)
    at ({\shiftwheelX+\rc*sin(72*\i)}, {\shiftwheelY+\rc*cos(72*\i)}) {};
}

\node[circle, fill=black, draw, scale=0.6] (hubcore) at (\shiftwheelX,\shiftwheelY) {};

\foreach \i in {0,...,3}{
  \pgfmathtruncatemacro{\j}{\i+1}
  \draw[line width=0.3mm] (w\i)--(w\j);
}
\draw[line width=0.3mm] (w4)--(w0);

\foreach \i in {0,...,4}{
  \draw[line width=0.3mm] (hubcore)--(w\i);
}
\end{tikzpicture}
\vspace{-0.8cm}
\includegraphics[width=\linewidth]{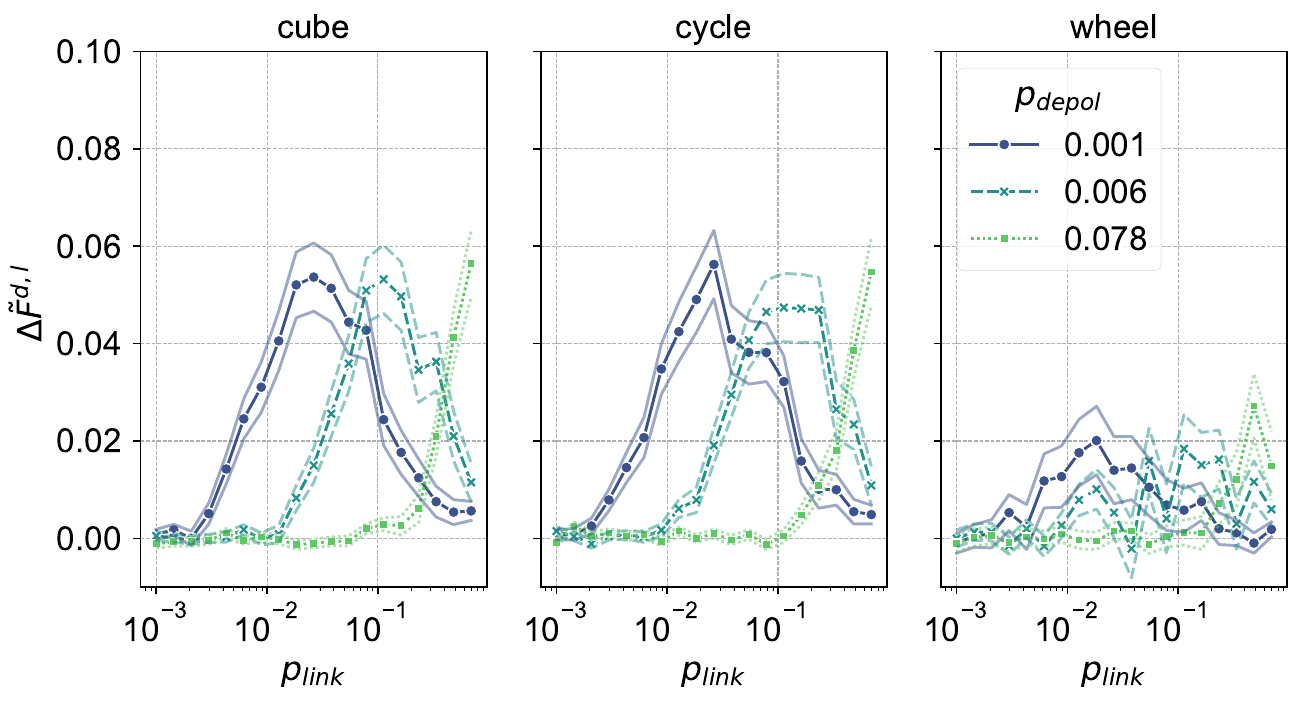}
\caption{Fidelity difference $\Delta \tilde{F}^{\mathrm{d,l}}$ between the general Piecemaker protocol utilizing LC equivalence and the Factory protocol for a set of example graph states reported in~\cite{adcock2020mapping}. The error bars indicate the standard error of the mean of $N=10^4$ simulation runs with a constant depolarizing probability $p_{\mathrm{depol}}=0.002$.}
\label{fig:adcockresults}
\end{figure}

We compare the fidelities of graph states produced by general Piecemaker using LC equivalence and Factory on three target graphs reported in~\cite{adcock2020mapping}: a 2-cube graph ($n=8$), a 8-cycle graph ($n=8$), and a 6-wheel graph ($n=6$). The results reflect the sizes of the MLCs~\cite{sharma2025minimizing}: cube and cycle graphs of size $n$ exhibit MLCs of size $n/2$, which is beneficial for Piecemaker, since it needs to wait for only half of the nodes to create links before memories can be released. The MLC for an $n$-wheel is comparatively larger with a size of $n-2$, resulting in fidelities close to those of Factory. In Figure \ref{fig:adcockresults} we thus see similar improvements for cube and cycle graphs with fidelity gains of approximately $\Delta F^* = 0.06$. The improvement for the wheel graph is less significant and stays below 0.04.

\section{Conclusion and Outlook}
In this study, we introduced protocols for the piece-wise distribution of GHZ states (GHZ Piecemaker) and general stabilizer states (general Piecemaker) to end nodes connected with a quantum switch. We compared the 
average fidelity of entangled states produced by our protocols to those achievable by a protocol introduced in~\cite{avis2023analysis}---the so-called Factory protocol. 
Our simulations showed that our protocols are generally able to achieve higher average fidelities. The GHZ Piecemaker achieved the largest average fidelity improvement of $0.2$ over Factory, where we observed up to a 45\%  relative decrease in infidelity. Our case study involving unequal end node-switch link lengths revealed that
the GHZ Piecemaker is less affected by inhomogeneity than the Factory, for the parameter regimes we explored. In addition to GHZ states, we provided a detailed recipe for generating and distributing general stabilizer states using our quantum networking-setup. Here, we used the important concept of LC equivalence that enables the switch to save storage time of individual Bell pairs in memory. Overall, the proposed schemes to generate general stabilizer states show promising results in considered parameter ranges, and improvement over the prior state of the art in terms of final quality of the generated states.

However, the studied protocols operate under various simplified assumptions and therefore form an initial contribution towards a broader understanding of stabilizer state distribution using quantum switches. Future goals include finding ways to further improve these protocols, as well as to evaluate them under more realistic hardware constraints. For example, a more restrictive assumption would be to introduce cutoff times, that is, to discard Bell pairs when they have undergone too much memory decoherence\cite{li2020efficient}. Furthermore, significant portions of the multipartite entanglement generation process leave many quantum memories idle. These resources could instead be used to create additional Bell pairs for the purpose of i) generating target states in a concurrent manner (which would increase the overall rate) or to ii) perform entanglement distillation~\cite{kalb2017entanglement, jansen2022enumerating, goodenough2023near, gu2025constant} which can lead to a higher-quality state, or a balanced combination of both.

\section{Methods}
\subsection{Simulation implementation and error model} \label{methods:simulation}
We simulate Piecemaker and Factory protocols as well as the considered quantum networking model using \textit{QuantumSavory.jl}, a multi-formalism simulator for noisy quantum communication and computation in Julia~\cite{quantumsavoryjl}. Note that the Factory \cite{avis2023analysis} was initially implemented in \textit{NetSquid}~\cite{coopmans2021netsquid}, a quantum networking simulator in Python. To ensure consistent comparison between Factory and Piecemaker as well as reproducability of results, using the same programming language environment is vital. We therefore replicated the Factory node in \textit{QuantumSavory.jl}.

We use the stabilizer formalism to represent quantum states. Stabilizer expressions and associated functionality are available in \textit{QuantumClifford.jl}. We model Markovian single-qubit noise over a time step $\Delta t = 1$ ms by sampling a Pauli error and applying it in place to the stabilizer tableau. Therefore, in each time step we draw an  operator $\sigma \in \{I, X, Y, Z\}$ and apply it to an addressed qubit $v$ with probability
$$
\Pr[\sigma = X_v] = \Pr[\sigma = Y_v] = \Pr[\sigma = Z_v] = \frac{p_\mathrm{depol}}{4}, 
$$
and
$$
\Pr[\sigma = I_v] = 1 - \frac{3p_\mathrm{depol}}{4},
$$ where $p_\mathrm{depol}$ is the depolarizing probability (see (\ref{eq:depol-prob})). Note that depolarizing noise is fundamentally a Poisson process (in continuous time), which we however model as a geometric process using fixed-size discrete time steps.
Because a single-qubit Clifford gate corresponds to a constant-time column swap and phase update in the tableau, the entire noise step is $\mathcal{O}(1)$ in both memory and time, enabling large-scale errors to be propagated under Pauli noise in linear time in the number of qubits.

\subsection{Generation of MLCs and associated graphs} \label{sec:genUG}

We detail here our method for finding the (minimal) local covers of a given graph $G$, which is based on the tools developed in~\cite{sharma2025minimizing}. Specifically, they construct an integer linear program that takes as input a graph $G$ on a vertex set $V$ and weights for each edge, i.e.~$W:\binom{V}{2}\rightarrow \mathbb{R}$. The program returns a locally equivalent graph $ G'$ that minimizes $\sum_{e \in E} W(e)$, the weighted sum over the edge set $E$ of $G'$.

Note that the above observation was also made in~\cite{sharma2025minimizing}, since the above problem is equivalent to determining whether the edgeless graph on the complement of $S$ is a vertex-minor of $G$~\cite{dahlberg2020transform}. One benefit of the tools from~\cite{sharma2025minimizing}
is the fact that an LC-equivalent graph $G'$ can be found such that the total number of edges is minimized, under the constraint that there are no edges internal to $\bar{S}$.

\section{Data and code availability}
The complete dataset analyzed in this study is archived in\cite{Prielinger2025Data}, which also contains the scripts used to reproduce all figures in the manuscript. The simulation code is part of the same data archive as well as the open source package \emph{QuantumSavory.jl}~\cite{quantumsavoryjl}.

\section{Acknowledgements}
This work was funded by the Army Research Office
(ARO) MURI on Quantum Network Science under grant
number W911NF2110325 and QuTech NWO funding 2020–2024 Part I ‘Fundamental Research’, Project Number 601.QT.001-1, financed by the Dutch Research Council (NWO). We further acknowledge support from
 NWO QSC grant BGR2 17.269. In addition, we acknowledge the following NSF grants: 2522101, 2402861, 2346089, 1941583.

 \section{Author contribution statement}
KG and DT conceived the Piecemaker protocol, and theoretical methodology. KG and GV supervised the project. KG conducted the design and numerical implementation of the MLCs discovery algorithm. LP conducted the numerical implementation of the time-dependent networking model with critical contributions by GA and SK. GV and GA conceived the analysis in the inhomogeneous setting. LP carried out the experiments, collection of data, interpretation of results and visualization. LP and KG wrote the manuscript and DT, GV, GA, SK provided critical revisions to the latter. All authors approved the final version of the manuscript.

\bibliographystyle{IEEEtran}

\end{document}